\documentclass[prx, aps, reprint, superscriptaddress, floatfix, longbibliography]{revtex4-2}
\usepackage{amsmath}
\usepackage{physics}
\usepackage{graphicx}
\usepackage{amssymb}
\usepackage{amsthm}
\usepackage{mathtools}
\usepackage[export]{adjustbox}
\usepackage[usenames,dvipsnames]{xcolor}
\definecolor{myblue}{rgb}{0,0,1}
\usepackage[breaklinks=true,colorlinks=true,linkcolor=myblue,citecolor=myblue]{hyperref}
\usepackage{tikz}

\newcommand{\I}{\mathrm{i}}
\newcommand{\matele}[3]{\langle #1|#2|#3\rangle}
\newcommand{\mcG}{\mathcal{G}}

\newcommand{\omax}{\omega_{\max}}

\newcommand{\OO}[1]{\mathcal{O}(#1)}
\newcommand{\paren}[1]{\left( #1 \right)}

\newcommand{\wh}[1]{\widehat{#1}}
\newcommand{\wt}[1]{\widetilde{#1}}

\renewcommand{\Re}{\operatorname{Re}}
\renewcommand{\Im}{\operatorname{Im}}

\makeatletter
\newcommand{\ostar}{\mathbin{\mathpalette\make@circled\star}}
\newcommand{\make@circled}[2]{%
  \ooalign{$\m@th#1\smallbigcirc{#1}$\cr\hidewidth$\m@th#1#2$\hidewidth\cr}%
}
\newcommand{\smallbigcirc}[1]{%
  \vcenter{\hbox{\scalebox{0.77778}{$\m@th#1\bigcirc$}}}%
}
\makeatother

\newtheorem{remark}{Remark}

\newcommand{\CCQ}{Center for Computational Quantum Physics, Flatiron Institute, 162 5th Avenue, New York, NY 10010, USA}
\newcommand{\CCM}{Center for Computational Mathematics, Flatiron Institute, 162 5th Avenue, New York, NY 10010, USA}

\begin{document}

\title{Decomposing imaginary time Feynman diagrams using separable
basis functions: Anderson impurity model strong coupling expansion}

\author{Jason Kaye}
\affiliation{\CCQ}
\affiliation{\CCM}

\author{Zhen Huang}
\affiliation{Department of Mathematics, University of California, Berkeley, CA 94720, USA}
\affiliation{\CCQ}

\author{Hugo U. R. Strand}
\affiliation{School of Science and Technology, Örebro University, SE-701 82 Örebro, Sweden}
\affiliation{Institute for Molecules and Materials, Radboud University, 6525 AJ Nijmegen, the Netherlands}

\author{Denis Gole\v z}
\affiliation{Jožef Stefan Institute, Jamova 39, SI-1000, Ljubljana, Slovenia}
\affiliation{Faculty of Mathematics and Physics, University of Ljubljana, Jadranska 19, 1000 Ljubljana, Slovenia}

\begin{abstract}
We present a deterministic algorithm for the efficient evaluation of imaginary time
diagrams based on the recently introduced discrete Lehmann
representation (DLR) of imaginary time Green's functions.
In addition to the efficient discretization of diagrammatic integrals afforded
by its approximation properties, the DLR basis is separable in imaginary time,
allowing us to decompose diagrams into linear combinations of nested sequences
of one-dimensional products and convolutions.
Focusing on the strong coupling bold-line expansion of generalized Anderson
impurity models, we show that our strategy reduces the computational complexity
of evaluating an $M$th-order diagram at inverse temperature $\beta$ and spectral width $\omax$ from
$\OO{(\beta \omax)^{2M-1}}$ for a direct quadrature to $\OO{M (\log (\beta \omax))^{M+1}}$, with
controllable high-order accuracy.
We benchmark our algorithm using third-order expansions for multi-band impurity problems with off-diagonal
hybridization and spin-orbit coupling, presenting
comparisons with exact diagonalization and quantum Monte Carlo approaches. In particular, we perform a self-consistent dynamical mean-field theory calculation for a three-band Hubbard model with strong spin-orbit coupling representing a minimal model of Ca$_2$RuO$_4$, demonstrating the promise 
of the method for modeling realistic strongly correlated multi-band materials.
For both strong and weak coupling expansions of low and intermediate order, in which diagrams can
be enumerated, our method provides an efficient, straightforward, and
robust black-box evaluation procedure.
In this sense, it fills a gap between diagrammatic approximations of the lowest
order, which are simple and inexpensive but inaccurate, and
those based on Monte Carlo sampling of high-order diagrams.
\end{abstract}

\maketitle

\section{Introduction}

Feynman diagram expansions are a standard computational tool in quantum many-body physics, both in condensed matter and quantum chemistry~\cite{georges1996,gull2011,houcke2010,onida2002}.
Given a Hamiltonian, one expands around an exactly solvable limit, such as the non-interacting (or atomic) limit, and interaction (or atom-atom coupling) corrections are captured by summing diagrams up to some order.
Directly-evaluated low-order expansions, like Hartree-Fock and Hedin's GW method, are routinely used in chemistry and solid-state physics first-principles calculations~\cite{onida2002,hedin1965,golze2019,deslippe2012berkeleygw}.
Similarly, the first-order bold expansion about the atomic limit, also called
the non-crossing approximation (NCA), is widely used
for quantum impurity problems ~\cite{costi1996,kroha1998,keiter1971,grewe1981,eckstein2010nonequilibrium,haule2001}.
While such low-order expansions are simple, inexpensive, and reliable for systems close to the exactly solvable limit, they are inadequate in the non-perturbative regime.
In certain cases, such as the description of Kondo resonances in impurity problems, including diagrams of even slightly higher order is required for the correct recovery of physical observables~\cite{pruschke1989, gull2010, haule2001,haule2023}.
However, direct evaluation of high-order expansions requires high-dimensional quadrature, rendering it impractical beyond even the first few orders.

We illustrate the state of the field for the example of the bold hybridization
expansions of the Anderson impurity model, which we focus on in this work. Here
the first-order NCA, which requires no integration, is used
routinely~\cite{georges1996,haule2007,eckstein2010nonequilibrium,golez2015,bittner2018},
as is the second-order one-crossing approximation (OCA), which requires
only two-dimensional
integration~\cite{de2019,eckstein2010nonequilibrium,pruschke1989,vildosola2015,korytar2011,strand2017,golez2019multi}.
At third-order and beyond, we are aware of only a
few studies due to the rapidly growing cost of direct
quadrature~\cite{haule2010,eckstein2010nonequilibrium,haule2023}, though expansions at this order describe the physics around the Mott transition remarkably well. A notable recent
exception involves a combination of
a three-point vertex computed by
direct quadrature,
and Monte Carlo sampling of the four-point vertex~\cite{kim2022,kim2023}. Thus,
although direct evaluation methods are simple, robust, and can be made high-order accurate with respect to quadrature error (see Sec.~\ref{sec:equad}), they have thus far been almost entirely restricted to the
lowest-order diagrammatic expansions, limiting their usefulness in addressing
challenging, material-realistic models.

In the opposite regime of very high-order expansions, diagrammatic Monte
Carlo methods for sampling over
diagram orders, topologies, and integrals have led to enormous success in
producing accurate results for sophisticated, strongly correlated systems, including
quantum impurity problems~\cite{rubtsov2005,gull2011,werner2006,werner2006a,Haule:2007ys} in
combination with dynamical mean-field theory (DMFT)~\cite{georges1996,Kotliar:2006aa}, polaron problems~\cite{prokof2007,hahn2018,mishchenko2000} and lattice Hubbard problems~\cite{rossi2017,simkovic2019,mountenet2018,schafer2021}. However, such approaches are computationally intensive, slowly converging (at the half-order Monte Carlo or first-order quasi-Monte Carlo~\cite{macek2020, strand2023inchworm} rate), and in many cases lack
robustness due to the sign problem \cite{gull2011}. In DMFT applications, the sign problem
has prevented the application of Monte Carlo-based methods to a large class of materials, such as multi-band systems with off-diagonal hybridization~\cite{eidelstein2020}, e.g.
spin-orbit coupled 4d and 5d electron systems. For several prominent correlated materials the sign problem has been mitigated by employing a basis transformation within the interaction expansion~\cite{zhang2017, zhang2016}, but this approach is limited to rather high temperatures. Another approach, the inchworm Monte Carlo formulation of the strong coupling expansion~\cite{eidelstein2020,cohen2015taming}, has been shown to mitigate the sign problem in minimal impurity models, and its range of applicability is being actively explored.
A promising recent development, the tensor train diagrammatics method, uses tensor cross interpolation (TCI) rather than Monte Carlo sampling, and was used to compute high-order bare expansions of the Anderson impurity problem, both for the interaction~\cite{fernandez22} and hybridization~\cite{erpenbeck2023} expansions, without a sign problem and with convergence rates significantly faster than Monte Carlo methods. This technique is related
in several ways to the algorithm presented here, but at present it has been used
primarily for high-order bare expansion diagrams, and since it relies on a specific
underlying compressibility structure of the integrand, its range of applicability
is not yet well-understood.

An opportunity therefore exists for the development of diagram evaluation
techniques which maintain the simplicity and robustness of direct methods at the
lowest orders while extending their range of applicability at least to intermediate
orders. Indeed, practitioners typically switch to diagrammatic Monte Carlo
methods even when the required expansion order is only slightly beyond the reach
of direct methods, due to the lack of practical alternatives~\cite{gull2010,kim2023}. By exploiting the specific structure of imaginary time diagrams, we obtain a
method which aims to make such calculations routine, deterministic, and
high-order accurate.  It relies on the recently-introduced discrete Lehmann
representation (DLR), which provides a compact basis of exponentials in which to
expand arbitrary single-particle imaginary time Green's functions, and related
quantities, with high-order accuracy \cite{kaye22_dlr}. Beyond the favorable
discretization properties of the DLR, we show that the separability of the DLR
basis functions in their imaginary time argument can be used to decompose
diagrams into linear combinations of nested sequences of products and
convolutions. These products and convolutions are then computed efficiently in
the DLR basis. Whereas the cost of direct evaluation methods for an $M$th-order diagrams scales with the inverse temperature $\beta$ and spectral width $\omax$ as
$\OO{(\beta \omax)^{2M-1}}$, our proposed method scales as $\OO{M (\log(\beta \omax))^{M+1}}$. 
The method is trivially parallelizable over the large number of diagrams appearing in typical diagrammatic calculations.

We implement our algorithm for the strong coupling expansion up to third-order, and benchmark it on several challenging Anderson impurity problems. We observe rapid order-by-order convergence within the Mott insulating regime for
systems with off-diagonal hybridization and/or strong local spin-orbit coupling. We also solve a minimal model for the strongly correlated calcium ruthenate Ca$_2$RuO$_4$
within DMFT~\cite{sutter2017,han2018,liebsch2007,georgescu2022,hao2020}. The significant spin-orbit coupling in this material makes it a challenging problem for Monte Carlo-based methods~\cite{zhang2017}, but we show that the third-order solution is highly accurate.
A general implementation of our approach beyond third-order diagrams and to weak coupling expansions is straightforward,
requiring only technical effort, and our formalism indicates a clear path towards extension 
to systems beyond impurity problems, like molecular or extended systems.
The main idea of our algorithm---separation of variables using sum-of-exponentials approximations---may be applicable to
higher order quantum many-body expansions comprising higher dimensional correlators and kernels,
such as the triangular vertex functions in the Hedin equations \cite{hedin1965, Aryasetiawan:2008aa, kim2022, kim2023}, and the two-particle objects appearing in the Bethe-Salpeter equation \cite{PhysRevLett.69.168, georges1996}.
Furthermore, our approach is likely complementary to other methods, such as TCI, which aim to address the exponential-scaling bottleneck of high-order diagrammatic calculations, providing a new ingredient in the design of algorithms based on these tools.
This work therefore represents a promising proof of concept, demonstrated on several challenging calculations, for a fundamental new tool in diagrammatic calculations.

\section{Overview of the method} \label{sec:overview}

The main idea of our algorithm can be demonstrated with a simple example whose structure is typical of imaginary time Feynman
diagrams. Consider the integral
\[I(\tau) = \int_0^\tau d\tau' \, \int_0^{\tau'} d\tau'' \, G_1(\tau-\tau'') \, G_2(\tau-\tau') \,
G_3(\tau'-\tau''),\]
for scalar or matrix-valued $G_i$ and $\tau \in [0, \beta]$. Since the factor $G_1$ couples the $\tau$ and $\tau''$ variables, $I(\tau)$
must be calculated as a double integral.
However, if we can make a low-rank approximation
\begin{equation} \label{eq:sep}
G_1(\tau - \tau'') \approx \sum_{l=1}^r G_{1l} \, \varphi_l(\tau)
\psi_l(\tau'')
\end{equation}
with $r$ small, for some scalar-valued $\varphi_l$ and $\psi_l$, then
we can separate variables:
\begin{equation*}
  \begin{multlined}
  I(\tau) \approx \sum_{l=1}^r G_{1l} \, \varphi_l(\tau) \int_0^\tau d\tau' \, G_2(\tau-\tau')
\\ \times \int_0^{\tau'} d\tau'' \, G_3(\tau'-\tau'') \, \psi_l(\tau'').
  \end{multlined}
\end{equation*}
$I(\tau)$ can then be computed from $r$ sequences of products and nested convolutions,
which may be less expensive than computing a full double integral for each
$\tau$, particularly if the $G_i$ and the product and convolution operations can be discretized efficiently.
In the case of imaginary time quantities, the DLR basis \cite{kaye22_dlr} provides an
efficient discretization with the separability property \eqref{eq:sep}.
The number of basis functions scales as $r = \OO{\log(\beta \omax) \log(1/\varepsilon)}$,
with $\varepsilon$ a user-specified accuracy, for any $G_1$.
We will apply this separation of variables approach to certain hybridization functions appearing in the strong coupling expansion, and compute the resulting products and convolutions in the DLR basis. We also demonstrate its application to the weak coupling expansion in Appendix~\ref{AppSec:Weak}.

\begin{remark}
  In the applied and computational mathematics
literature, separability in
sum-of-exponentials approximations has been used to obtain
fast algorithms for applying nonlocal integral operators in a variety of
settings~\cite{beylkin05, beylkin10, zhang21, gao22}, including fast history integration and compression in Volterra integral equations~\cite{greengard00,jiang15,hoskins23} and nonlocal transparent boundary conditions~\cite{alpert02,jiang04,jiang08}, diagonal translation operators in the fast multipole method~\cite{greengard97,cheng99,hrycak98}, the fast Gauss transform~\cite{jiang21}, the periodic fast multipole method~\cite{pei23}, and others~\cite{gimbutas20,barnett23}. In these
applications, one typically considers an integral transform $I(x) = \int dx' \,
K(x-x') f(x')$ for a kernel $K$ which is known a priori, and uses a sum-of-exponentials approximation of $K$ to separate ``source'' (internal, or integration) variables $x'$ from
``target'' (external) variables $x$. By contrast, Feynman diagrams involve higher-dimensional integrals
connecting a priori unknown functions entangled via many internal and external
variables.
\end{remark}

\section{Background: diagrammatic methods and numerical tools}

\subsection{Strong coupling hybridization expansion}

Quantum impurity problems are zero-dimensional interacting quantum many-body systems in contact with a general bath environment.
The local part of the impurity Hamiltonian
  $H_\text{loc}$ can have arbitrary quadratic terms $\epsilon_{\kappa\lambda}$
  and quartic terms $U_{\kappa \lambda \mu \nu}$:
\begin{equation}
\label{Eq:Ham}
  H_\text{loc} =
  \sum_{\kappa, \lambda = 1}^{n} \epsilon_{\kappa \lambda} c^\dagger_\kappa c_\lambda +
  \sum_{\kappa, \lambda, \mu, \nu = 1}^{n} U_{\kappa \lambda \mu \nu} c^\dagger_\kappa c^\dagger_\lambda c_\mu c_\nu
  \, .
\end{equation}
Here $c^\dagger_\lambda$ is the creation operator for a fermion in the impurity state $\lambda$ and $n$ is the number of impurity states. 
The full quantum impurity problem, including the coupling to the bath, can be described in terms of the action
\begin{equation}
\label{Eq:Act_hyb}
  \begin{multlined}
  \mathcal{S} =
  \int_0^\beta d\tau \, H_\text{loc}[c, c^\dagger]
  \\ + \sum_{\kappa, \lambda = 1}^{n} \int_0^\beta d\tau \, \int_0^\beta d\tau' \,
  c^\dagger_\lambda(\tau) \Delta_{\lambda \kappa}(\tau - \tau') c_\kappa(\tau').
  \end{multlined}
\end{equation}
The hybridization function $\Delta_{\lambda\kappa}(\tau - \tau')$ describes the propagation of a fermion in impurity state $\kappa$
 entering the bath at time $\tau'$ and returning to the impurity state $\lambda$ 
 at time $\tau$. $\Delta_{\lambda \kappa}$ is a scalar-valued function for each fixed $\kappa$ and $\lambda$.

The properties of the impurity problem can be characterized in terms of static expectation values and dynamical response functions. We focus here on the single-particle Green's function $G_{\lambda\kappa}(\tau - \tau') = - \langle \mathcal{T} c_\lambda(\tau) c^\dagger_\kappa(\tau') \rangle$, which describes the temporal correlation between the addition of a fermion to the impurity in state $\lambda$ and the removal of a fermion in state $\kappa$.
In the non-interacting limit $U_{\kappa\lambda\mu\nu}=0$, the Green's function can be determined analytically, and for non-zero interactions one can carry out an expansion in the interaction parameter, called the interaction expansion.
However, for many strongly correlated systems this becomes infeasible, requiring high expansion orders~\cite{gull2007performance, tsuji2013nonequilibrium}.  
For sufficiently strong interactions, the series diverges with perturbation order, requiring tailored resummations derived from conformal transformations~\cite{bertrand2019}.

In the limit of an impurity decoupled from the bath (zero hybridization $\Delta_{\lambda \kappa} = 0$), we can directly diagonalize $H_\text{loc}$ since there are a finite number of local many-body states. 
This is the starting point of the strong coupling expansion, which is in essence a perturbative expansion in the hybridization function. We refer to Ref.~\onlinecite{eckstein2010nonequilibrium} for a detailed description of this approach, and briefly summarize its main characteristics here.

To enable the hybridization expansion, the impurity action $\mathcal{S}$ is rewritten by introducing a pseudo-particle $p^\dagger_k \ket{0}$ for each impurity many-body state $\ket{k}$, making the local Hamiltonian quadratic and the hybridization quartic in the pseudo-particle space. The resulting action is given by 
\begin{equation}
  \begin{multlined}
    \mathcal{S} =
    \sum_{j, k = 1}^{N}
    \int_0^\beta d\tau \, p^\dagger_j(\tau)
    \matele{j}{H_\text{loc}}{k}  
    p_k(\tau)
    \\ +
    \sum_{j,k,j',k'=1}^{N}
    \sum_{\kappa, \lambda=1}^{n} \int_0^\beta d\tau \, \int_0^\beta d\tau' \,
    p^\dagger_j(\tau) p_k(\tau)
    F^\dagger_{\lambda j k} \\
    \times \Delta_{\lambda \kappa}(\tau - \tau')
    F_{\kappa k' j'}
    p^\dagger_{k'}(\tau') p_{j'}(\tau'),
  \end{multlined}
\end{equation}
where $N = 2^{n}$ is the number of local many-body states and $F_{\kappa j k}=\matele{j}{c_\kappa}{k}$.
This action can be expanded in the quartic hybridization term.
The pseudo-particle Green's function $\mathcal{G}_{jk}(\tau-\tau') = - \langle \mathcal{T} p_j(\tau) p^\dagger_k(\tau') \rangle$ 
satisfies the Dyson equation
\begin{equation}
  \mathcal{G}
  =
  g + g \ast \Sigma \ast \mathcal{G}
  \, ,
  \label{eq:ppdyson}
\end{equation}
where $\Sigma$ is the pseudo-particle self-energy, $g_{jk}(\tau-\tau') = -\langle \mathcal{T} p_j(\tau) p_k^\dagger(\tau') \rangle_{\Delta=0}$
 is the non-interacting ($\Delta=0$) pseudo-particle Green's function, and $\ast$ denotes the time-ordered convolution
\[(a*b)_{jk}(\tau)=\sum_{l=1}^N \int_0^{\tau} d\tau' \, a_{jl}(\tau - \tau') b_{lk}(\tau').\]

The pseudo-particle self-energy $\Sigma$ contains the following sequence of diagrams:
\begin{equation}
  \tikz[baseline=(current bounding box.center)]
  \node[] {\includegraphics[scale=1]{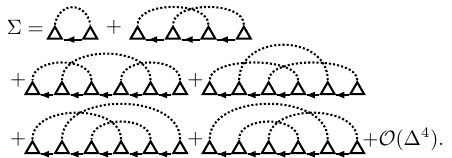}};
  \label{eq:sigma}
\end{equation}
Solid lines correspond to the pseudo-particle Green's function:
\begin{equation*}
  \tikz[baseline=(current bounding box.center)]
  \node[] {\includegraphics[scale=1]{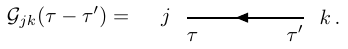}};
\end{equation*}
Each undirected dotted line 
\begin{equation}
  \tikz[baseline=(current bounding box.center)]
  \node[] {\includegraphics[scale=1]{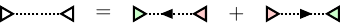}};
  \, 
  \label{eq:dir_sum}
\end{equation}
corresponds to a sum over forward and backward propagation of the hybridization function interaction. A forward hybridization function interaction is represented by
\begin{equation}
  \tikz[baseline=(current bounding box.center)]
  \node[] {\includegraphics[scale=1]{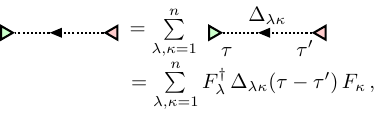}};
  \label{eq:hybr_sum_forw}
\end{equation}
which in turn contains a sum over hybridization functions $\Delta_{\lambda \kappa}$. 
$F_\kappa$, represented by a red triangle, is an $N \times N$ matrix with entries $F_{\kappa m n}$, and similarly for $F_\lambda^\dagger$, which is represented by a green triangle.
A backward interaction is given by
\begin{equation}
  \tikz[baseline=(current bounding box.center)]
  \node[] {\includegraphics[scale=1]{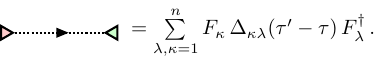}};
  \label{eq:hybr_sum_back}
\end{equation}
The order $M$ 
of each self-energy diagram in \eqref{eq:sigma} is given by the number of hybridization interactions (dotted lines) propagating either forward or backward.
Each diagram is composed of a backbone of forward-propagating pseudo-particle Green's functions $\mathcal{G}$ (solid lines)
connected by vertices associated with one end of a hybridization line. 
Each vertex represents an insertion of a matrix $F_\lambda^\dagger$ or $F_\kappa$ at a given time $\tau_i$.
The internal times $\tau_1, \ldots, \tau_{2M-2}$ are integrated over in the domain $\tau_1 \le \cdots \leq \tau_{2M-2} \leq \tau$.
The prefactor of each diagram is $(-1)^{s+f+M}$, where $s$ is the number of crossing hybridization lines and $f$ is the number of backward-propagating hybridization lines. We give specific examples with mathematical expressions in the next subsection.

The single-particle Green's function $G(\tau)$ can be recovered from the pseudo-particle Green's function $\mathcal{G}$ using the circular diagram series 
\begin{equation}
  \tikz[baseline=(current bounding box.center)]
  \node[] {\includegraphics[scale=1]{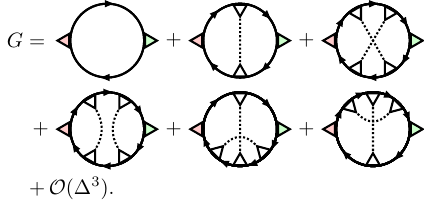}};
  \label{eq:spgf}
\end{equation}
We again describe the construction of these diagrams, and give examples in the next subsection.
The diagram order $M$ is in this case one more than the number of hybridization interactions (dotted lines). Each diagram consists of a closed loop of pseudo-particle propagators $\mathcal{G}$ with two extra operator vertices $F_\lambda$ and $F^\dagger_\kappa$ inserted at the times $\tau$ and $0$ (red and green triangles, respectively). Here, $\lambda$ and $\kappa$ are the single-particle state indices of the single-particle Green's function $G_{\lambda\kappa}(\tau - 0)$. The hybridization interactions and associated vertices have the same structure as in the self-energy diagrams.
For $M > 1$ the internal times $\tau_1, \ldots, \tau_{2M-2}$ are integrated over in the domain $0 \leq \tau_1 \leq \cdots \leq \tau \leq \cdots \leq \tau_{2M-2} \leq \beta$. The number of internal times on the intervals $[0, \tau]$ and $[\tau, \beta]$, respectively, varies from one diagram to another. A trace is taken over the $N$-dimensional pseudo-particle state indices.
The sign of a diagram is determined by first inserting a hybridization line between the two external times $\tau$ and $0$, and then cutting an arbitrary pseudo-particle propagator. The prefactor is obtained from the modified diagram as $(-1)^{s+f+M}$.

The steps required to compute the single-particle Green's function $G$ can be summarized as follows: (i) the pseudo-particle Green's function $\mathcal{G}$ and self-energy $\Sigma$ are determined self-consistently by solving the Dyson equation \eqref{eq:ppdyson}, using the self-energy expansion \eqref{eq:sigma}, and (ii) $G$ is then obtained by evaluating the diagrams in \eqref{eq:spgf}.

\subsection{Examples of diagrams}
\label{sec:example}

Both the self-energy diagrams and the circular diagrams for the single-particle Green's function beyond first-order take the form of multidimensional integrals in imaginary time. We present typical examples for each case to elucidate their common structure.

\subsubsection{Pseudo-particle self-energy diagrams}

The approximation of the pseudo-particle self-energy that includes only first-order diagrams, called the non-crossing approximation (NCA), requires multiplication of $N \times N$ matrices but no integration:
\begin{equation}
  \tikz[baseline, anchor=base]
  \node[] at (0, -1em) {\includegraphics[scale=1]{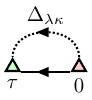}};
  =
  (-1)^{0+0+1}\Delta_{\lambda\kappa}(\tau - 0)
  F^\dagger_{\lambda} \, \mathcal{G}(\tau - 0) \, F_{\kappa}. 
\end{equation}
The complete first-order contribution to \eqref{eq:sigma} is given by
\begin{equation} \label{eq:ncacomplete}
  \tikz[baseline, anchor=base]
  \node[] at (0, -1em) {\includegraphics[scale=1]{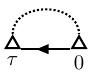}};
  =
  \sum_{\kappa, \lambda=1}^n
  \Bigg[
  \tikz[baseline, anchor=base]
  \node[] at (0, -1em) {\includegraphics[scale=1]{diagrams/sigma_ex1.pdf}};
  +
  \tikz[baseline, anchor=base]
  \node[] at (0, -1em) {\includegraphics[scale=1]{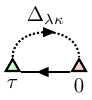}};
  \Bigg]
  \, ,
\end{equation}
i.e.\ the sum over all hybridization directions (see \eqref{eq:dir_sum})
and hybridization insertions $\Delta_{\lambda\kappa}$ (see
\eqref{eq:hybr_sum_forw} and \eqref{eq:hybr_sum_back}).

The second-order approximation to the self-energy is called the one-crossing approximation~(OCA), and contributing diagrams are given by double integrals:
\begin{equation} \label{eq:oca}
  \begin{multlined}
    \tikz[baseline, anchor=base]
    \node[] at (0, -1.5em) {\includegraphics[scale=1]{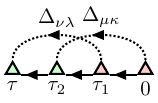}};
    =
    (-1)^{1+0+2}
    \\\int_{0}^{\tau} \!\! d\tau_2 \int_{0}^{\tau_2} \!\! d\tau_1 \,
    \Delta_{\nu\lambda}(\tau-\tau_1) \Delta_{\mu\kappa}(\tau_2)
    \\ \times
      F^\dagger_{\nu} \, \mathcal{G}(\tau - \tau_2) \,
        F^\dagger_{\mu} \, \mathcal{G}(\tau_2-\tau_1) \,
      F_{\lambda} \, \mathcal{G}(\tau_1) \, F_{\kappa}.
  \end{multlined}
\end{equation}
The complete second-order contribution to \eqref{eq:sigma} is obtained in a manner analogous to \eqref{eq:ncacomplete}.
We note that the factors corresponding to the forward-propagating backbone of impurity propagators have a repeated convolutional structure in the imaginary time variables. This structure is broken by the hybridization functions, which couple non-adjacent time variables.
All higher-order self-energy diagrams share this pattern.
For example, the diagrams comprising the first third-order contribution in \eqref{eq:sigma} are given by
\begin{widetext}
\begin{equation}
  \begin{multlined}
    \tikz[baseline, anchor=base]
    \node[] at (0, -1.5em) {\includegraphics[scale=1]{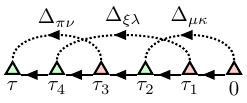}};
    =
    (-1)^{2+0+3}
    \int_{0}^{\tau} d\tau_4 \, \int_{0}^{\tau_4} d\tau_3 \, \int_{0}^{\tau_3} d\tau_2 \, \int_{0}^{\tau_2} d\tau_1 \,
    \Delta_{\pi\nu}(\tau-\tau_3) \Delta_{\xi\lambda}(\tau_4-\tau_1)\Delta_{\mu\kappa}(\tau_2)
    \\ \times
      F^\dagger_{\pi} \, \mathcal{G}(\tau\!-\!\tau_4) \,
      F^\dagger_{\xi} \, \mathcal{G}(\tau_4\!-\!\tau_3) \,
      F_{\nu} \, \mathcal{G}(\tau_3\!-\!\tau_2) \,
      F^\dagger_{\mu} \, \mathcal{G}(\tau_2\!-\!\tau_1) \,
      F_{\lambda} \, \mathcal{G}(\tau_1\!-\!0) \, F_{\kappa}.
  \end{multlined}
\end{equation}
\end{widetext}
Our strategy will be to reinstate the convolutional structure of the backbone by separating variables in the hybridization functions.

\subsubsection{Single-particle Green's function diagrams}

The diagrams for the single-particle Green's function $G_{\lambda\kappa}(\tau) = -\langle \mathcal{T} c_\lambda(\tau) c^\dagger_\kappa(0) \rangle$ contain two additional operators compared with the self-energy diagrams: $c_\lambda$ is inserted at time $\tau$, and $c^\dagger_\kappa$ is inserted at time $0$.
The first-order (NCA) diagrams in \eqref{eq:spgf} take the simple form
\begin{equation}
  \begin{multlined}
  \tikz[baseline, anchor=base]
  \node[] at (0, 0em) {\includegraphics[scale=1]{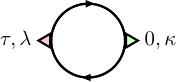}};
  \\=
  (-1)^{0+0+0}
  \text{Tr} \left[  \mathcal{G}(\beta-\tau) \, F_{\lambda} \, \mathcal{G}(\tau) \, F^\dagger_{\kappa} \right].
  \end{multlined}
\end{equation} \\
Since no hybridization function connects the times $0$ and $\tau$ in these diagrams, the indices $\kappa$ and $\lambda$ are included in the notation.
The second-order (OCA) diagrams contain a single hybridization insertion and two internal time integrals:
\begin{equation}
  \begin{multlined}
    \tikz[baseline, anchor=base]
    \node[] at (0, 0em) {\includegraphics[scale=1]{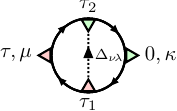}};
    \\ =
    (-1)^{1+0+1}
    \int_{\tau}^{\beta} d\tau_2 \, \int_{0}^{\tau} d\tau_1 \,
    \Delta_{\nu \lambda}(\tau_2-\tau_1) \\
    \times \text{Tr} \left[
      \mathcal{G}(\beta-\tau_2) \, F^\dagger_{\nu} \,
      \mathcal{G}(\tau_2-\tau) \, F_{\mu} \right.
      \\ \left. \times
      \mathcal{G}(\tau-\tau_1) \,  F_{\lambda} \,
      \mathcal{G}(\tau_1) \,  F^\dagger_{\kappa}
      \right]
      .  
  \end{multlined}
  \label{eq:spgf2}
\end{equation}
The third-order diagrams contain two hybridization insertions and four internal time integrals, e.g.,
\begin{widetext}
\begin{equation}
  \begin{multlined}
    \tikz[baseline, anchor=base]
    \node[] at (0, -2.7em) {\includegraphics[scale=1]{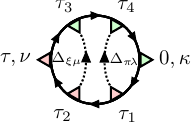}};
    =
    (-1)^{2+0+2}
    \int_{\tau}^{\beta} \!\! d\tau_4 \int_{\tau}^{\tau_4} \!\! d\tau_3
    \int_{0}^{\tau} \!\! d\tau_2 \int_{0}^{\tau_2} \!\! d\tau_1 \,
     \Delta_{\pi\lambda}(\tau_4\!-\!\tau_1) \Delta_{\xi\mu}(\tau_3\!-\!\tau_2)
    \\ \times
    \text{Tr} \left[
      \mathcal{G}(\beta\!-\!\tau_4) \, F^\dagger_{\pi} \,
      \mathcal{G}(\tau_4\!-\!\tau_3) \, F^\dagger_{\xi} \,
      \mathcal{G}(\tau_3\!-\!\tau) \, F_{\nu} \,
      \mathcal{G}(\tau\!-\!\tau_2) \, F_{\mu} \,
      \mathcal{G}(\tau_2\!-\!\tau_1) \, F_{\lambda} \,
      \mathcal{G}(\tau_1\!-\!0) \,  F^\dagger_{\kappa}
      \right]
    .
  \end{multlined}
\end{equation}
\end{widetext}
The backbone propagators again have a simple convolutional structure, now split into the two separable intervals $[0, \tau]$ and $[\tau, \beta]$, but the hybridization insertions again break this structure. 

\subsection{Evaluation by direct quadrature}
\label{sec:equad}

In order to establish a baseline for comparison with our approach, we describe a simple integration strategy based on equispaced quadrature rules which has been employed in the literature~\cite{haule2010,eckstein2010nonequilibrium, schuler20}. Since we focus on the evaluation of individual diagrams, for the remaining discussion we simplify notation by fixing the hybridization indices and
absorbing the matrices $F$, $F^{\dagger}$ into the Green's functions.
For example, we can write each OCA self-energy diagram \eqref{eq:oca} in the common form
\begin{equation} \label{eq:ocasimple}
  \begin{multlined}
  \Sigma(\tau) = \int_0^{\tau} d\tau_2 \, \int_0^{\tau_2} d\tau_1 \,
  \Delta_2(\tau-\tau_1) \, \Delta_1(\tau_2)  \\ \times \mcG_3(\tau-\tau_2) \,
  \mcG_2(\tau_2-\tau_1) \, \mcG_1(\tau_1),
  \end{multlined}
\end{equation}
where $\Sigma$ and the $\mcG_i$ are $N \times N$ matrix-valued and the $\Delta_i$ are
scalar-valued.

A simple approach is to
pre-evaluate all functions on an equispaced grid $\tau_j$ in imaginary time and discretize the integrals by the second-order accurate trapezoidal rule.
This reduces the double integral to
\[\Sigma(\tau_j) \approx \sideset{}{'}\sum_{k_2=0}^j \sideset{}{'}\sum_{k_1=0}^{k_2}
  \Delta_{2,j-k_1} \, \Delta_{1,k_2} \, \mcG_{3,j-k_2} \,
  \mcG_{2,k_2-k_1} \, \mcG_{1,k_1},\]
where we have used the notation
$\Delta_{i,j} = \Delta_i(\tau_j)$. The
prime on the sum indicates that its first and last terms
are multiplied by the trapezoidal rule weight $1/2$, unless the sum contains only one term, in
which case it is set to zero.
If $n_\tau$ grid points are used in each dimension,
then this method scales as $\OO{n_\tau^{2M-1}}$ ($2M-2$ internal time variables are integrated over for each $\tau$).
Furthermore, achieving convergence
in general requires taking $n_\tau = \OO{\beta \omax}$, with $\omax$ the maximum spectral width of all quantities
appearing in the integrand \cite{kaye22_dlr}. Defining the dimensionless constant $\Lambda = \beta \omax$, the scaling of this method is $\OO{\Lambda^{2M-1}}$.  

\begin{remark} \label{rem:highorder}
Although it does not improve the scaling with respect to $M$ or
$\Lambda$, the order of accuracy $p$---that is, the error convergence rate $n_\tau^{-p}$, given by $p=2$ for the trapezoidal rule---can be substantially improved at negligible additional cost using endpoint-corrected equispaced quadratures. For example, Gregory quadratures (yielding roughly $p \leq 10$), and more stable variants (yielding larger $p$) \cite{fornberg19,fornberg21}, increase the order of accuracy by reweighting a few endpoint values. For stability at very high-order accuracy, endpoint node locations must be modified, as in Alpert quadrature \cite{alpert99}, requiring high-order accurate on-the-fly evaluation of the integrand for a small subset of terms. To the authors' knowledge, such approaches have not yet been used in the literature for diagram evaluation, though Gregory quadratures up to order $p=6$ have been used in nonequilibrium Green's function calculations for real time integrals \cite{schuler20}.
Other possibilities include spectral methods like Gauss quadrature, yielding
spectral accuracy, or composite spectral methods, yielding arbitrarily
high-order accuracy, but these rules are not based on an underlying
equispaced grid and therefore require on-the-fly evaluation of the integrand on irregular grids.
\end{remark}

\subsection{Discrete Lehmann representation}

We give a short summary of the main properties of the DLR used by our algorithm. For a detailed
description of the DLR, we refer to Ref.~\onlinecite{kaye22_dlr}, and
to Ref.~\onlinecite{kaye22_libdlr} for another brief overview.
The DLR, like the closely related intermediate representation (IR) \cite{shinaoka17,chikano18}, is based on the spectral Lehmann
representation
\begin{equation} \label{eq:lehmann}
  G(\tau) = -\int_{-\infty}^\infty d\omega \, K(\tau,\omega) \rho(\omega)
\end{equation}
of imaginary time Green's functions. Here, $G$ is a Green's function, $\rho$ is
its integrable spectral function, and
\begin{equation} \label{eq:kernel}
  K(\tau,\omega) = \frac{e^{-\tau \omega}}{1 + e^{-\beta \omega}}
\end{equation}
is called the analytic continuation kernel. Given the support constraint $\rho(\omega) = 0$ for $\omega$ outside $[-\omax,\omax]$, and defining
$\Lambda = \beta \omax$ as above, one observes that the singular values of the
integral operator defining the representation \eqref{eq:lehmann} decay
super-exponentially. In particular, the $\varepsilon$-rank $r$---the number of
singular values larger than $\varepsilon$---is $\OO{\log(\Lambda)
\log(1/\varepsilon)}$. This implies that the image of the operator, which
contains all imaginary time Green's functions, can be characterized to accuracy
$\varepsilon$ by a basis of only $r = \OO{\log(\Lambda) \log(1/\varepsilon)}$
functions.

Taking these functions to be the left singular vectors of the
operator yields the orthogonal IR basis.
Alternatively, taking the functions to be $K(\tau,\omega_l)$ for $r$ carefully
chosen $\omega_l$ yields the non-orthogonal but explicit DLR basis.
In particular, it is shown in Ref.~\onlinecite{kaye22_dlr} that the DLR frequencies $\omega_l$ can be
selected, using rank-revealing pivoted Gram-Schmidt orthogonalization, to give the
DLR expansion
\begin{equation} \label{eq:dlrexp}
  G(\tau) \approx \sum_{l=1}^r K(\tau, \omega_l) \wh{g}_l
\end{equation}
accurate to $\varepsilon$, with $r$ possibly slightly larger than the
$\varepsilon$-rank of the operator, or the number of its singular values greater than $\varepsilon$. We emphasize that $r$ and the DLR
frequencies depend only on $\Lambda$ and $\varepsilon$, and not on
$G(\tau)$ itself; to $\varepsilon$ accuracy, the span of the DLR basis contains
all imaginary time Green's functions satisfying the user-specified cutoff $\Lambda$.

Using a similar pivoted Gram-Schmidt procedure, one can construct a set of $r$ DLR
interpolation nodes $\tau_k$ such that the DLR coefficient $\wh{g}_l$ can be
stably recovered from samples $G(\tau_k)$ by solving the $r \times r$ linear system
\[G(\tau_k) = \sum_{l=1}^r K(\tau_k, \omega_l) \wh{g}_l.\]
This is similar to the sparse sampling method \cite{li20}, typically used in
conjunction with the IR basis, which obtains stable interpolation grids from
the extrema of the highest-degree IR basis function.
Green's functions can then be represented by their values on this DLR grid, and
operations can be carried out using this
representation. For example, given Green's functions $F(\tau)$ and $G(\tau)$ represented
by their DLR grid samples $F(\tau_k)$ and $G(\tau_k)$, we can evaluate their product $H = F G$ on the
DLR grid, $H(\tau_k) = F(\tau_k) G(\tau_k)$. Then the DLR expansion of
$H$ can be obtained as described above. An efficient algorithm to compute the convolution $H(\tau) =
\int_0^\beta d\tau' \, F(\tau-\tau') G(\tau')$ or time-ordered convolution $H(\tau) =
\int_0^\tau d\tau' \, F(\tau-\tau') G(\tau')$ is described in Appendix
\ref{app:fastconv}. We note that our method assumes self-energies and hybridization
functions, as well as products and convolutions of DLR expansions, can be
represented accurately in the DLR basis, which has been observed to be the case
in many previous works~\cite{kaye22_dlr,sheng23,labollita23,kaye23_eqdyson,li20,yeh22,cai22,shinaoka22}.

\section{Efficient evaluation of imaginary time diagrams} \label{sec:algorithm}

Our algorithm improves the $\OO{\Lambda^{2M-1}}$
scaling of the standard equispaced integration method described in
Section \ref{sec:equad} to $\OO{(2M-2)r^{M+1}} =
\OO{(2M-2)(\log \Lambda )^{M+1}}$.
It exploits the separability of the analytic continuation kernel, and therefore the DLR basis functions:
\begin{equation} \label{eq:ksep}
  K(\tau-\tau',\omega) = \frac{K(\tau,\omega)
  K(\tau',-\omega)}{K(0,-\omega)}.
\end{equation}
Using \eqref{eq:ksep}, we can separate variables in the hybridization functions which break the convolutional structure of the backbone,
reducing diagrams to sums over nested products and convolutions. Each such operation can then be evaluated efficiently in the DLR basis, as described above.

We first demonstrate the technique using the OCA-type self-energy diagram \eqref{eq:ocasimple}. 
Replacing $\Delta_2$ by its DLR expansion
$\Delta_2(\tau) = \sum_{l=1}^r
K(\tau,\omega_l) \wh{\Delta}_{2l}$ and separating variables gives
\begin{equation} \label{eq:deltasep}
  \Delta_2(\tau-\tau_1) = \sum_{l=1}^r \frac{K(\tau,\omega_l)
K(\tau_1,-\omega_l)}{K(0,-\omega_l)} \wh{\Delta}_{2l},
\end{equation}
and inserting this expression into \eqref{eq:ocasimple} gives
\begin{equation} \label{eq:ocasep1}
  \begin{multlined}
  \Sigma(\tau) = \sum_{l=1}^r \frac{\wh{\Delta}_{2l}}{K(0,-\omega_l)} K(\tau,\omega_l) \\ \times \int_0^\tau
  d\tau_2 \, \mcG_3(\tau-\tau_2) \Delta_1(\tau_2) \\ \times \int_0^{\tau_2} d\tau_1 \,
  \mcG_2(\tau_2-\tau_1) \mcG_1(\tau_1) K(\tau_1,-\omega_l).
  \end{multlined}
\end{equation}
Each term of the sum now consists of a nested sequence of one-dimensional
products and convolutions, which can be evaluated by the following procedure:
(1) multiply $\mcG_1$ and $K(\cdot,-\omega_l)$,
(2) convolve by $\mcG_2$,
(3) multiply by $\Delta_1$,
(4) convolve by $\mcG_3$, and
(5) multiply by $\frac{\wh{\Delta}_{2l}}{K(0,-\omega_l)} K(\cdot,\omega_l)$.
Here, products can be taken pointwise on the DLR grid of
$r$ nodes, and convolutions can be computed at an $\OO{r^2}$ cost using the method
described in Appendix \ref{app:fastconv}. 

A final technical point on numerical stability must be addressed. Since
$1/K(0,-\omega_l) = 1+e^{\beta \omega_l}$, \eqref{eq:ocasep1} is
vulnerable to overflow if $\omega_l > 0$. In this case, we can rewrite
\eqref{eq:deltasep} using
\begin{equation}
  K(\tau-\tau',\omega) = \frac{K(\tau-\tau'',\omega)
K(\tau''-\tau',\omega)}{K(0,\omega)}
\end{equation}
in place of \eqref{eq:ksep} to obtain
\begin{equation}
  \begin{multlined}
    \Delta_2(\tau-\tau_1) = \sum_{\omega_l \leq 0} \frac{K(\tau,\omega_l)
K(\tau_1,-\omega_l)}{K(0,-\omega_l)} \wh{\Delta}_{2l} \\ + \sum_{\omega_l >
0} \frac{K(\tau-\tau_2,\omega_l)
K(\tau_2-\tau_1,\omega_l)}{K(0,\omega_l)} \wh{\Delta}_{2l}.
  \end{multlined}
\end{equation}
This gives a numerically stable replacement of \eqref{eq:ocasep1}:
\begin{equation} \label{eq:ocasep2}
  \begin{multlined} 
  \Sigma(\tau) = \sum_{\omega_l \leq 0} \frac{\wh{\Delta}_{2l}}{K_l^-(0)} K_l^+(\tau) \int_0^\tau
    d\tau_2 \, \mcG_3(\tau-\tau_2) \Delta_1(\tau_2) \\ \times \int_0^{\tau_2} d\tau_1 \,
    \mcG_2(\tau_2-\tau_1) (\mcG_1 K_l^-)(\tau_1) \\
    + \sum_{\omega_l > 0} \frac{\wh{\Delta}_{2l}}{K_l^+(0)} \int_0^\tau
    d\tau_2 \, (\mcG_3 K_l^+)(\tau-\tau_2) \Delta_1(\tau_2)
    \\ \times \int_0^{\tau_2} d\tau_1 \,
    (\mcG_2 K_l^+)(\tau_2-\tau_1) \mcG_1(\tau_1).
  \end{multlined}
\end{equation}
Here, we have introduced the notation
\begin{equation}
  K_l^\pm(\tau) \equiv K(\tau,\pm \omega_l)
\end{equation}
and
\begin{equation}
  (\mcG_i K_l^\pm)(\tau) \equiv \mcG_i(\tau) K(\tau,\pm\omega_l).
\end{equation}
The procedure to evaluate the terms with $\omega_l \leq 0$ is the same as above,
but for those with $\omega_l > 0$ it is slightly modified:
(1) multiply $\mcG_2$ and $K_l^+$,
(2) convolve the result with $\mcG_1$,
(3) multiply by $\Delta_1$,
(4) multiply $\mcG_3$ and $K_l^+$,
(5) convolve with the previous result, and
(6) multiply by $\wh{\Delta}_{2l}/K_l^+(0)$.

\subsection{General procedure} \label{sec:procedure}

This idea may be generalized to arbitrary $M$th-order pseudo-particle self-energy and single-particle Green's function diagrams, containing
internal time integration variables $\tau_1,\ldots,\tau_{2M-2}$, using the following procedure.
Let $\Delta$ correspond to a hybridization line which does not connect to time zero, with DLR coefficients $\wh{\Delta}_l$. Order all imaginary time
  variables, including the variable $\tau$, as $\tau'_1 \leq \tau'_2 \leq \cdots \leq \tau'_{2M-1}$.
  Thus, for the self-energy diagrams, we have $\tau_i = \tau_i'$ for $i=1,\ldots,2M-2$,
  and $\tau_{2M-1}' = \tau$. For the Green's function diagrams, we have some $k <2M-1$ such that
  $\tau_i' = \tau_i$ for $i = 1,\ldots,k-1$, $\tau_k' = \tau$, and $\tau_i' =
  \tau_{i-1}$ for $i = k+1,\ldots,2M-1$. Replace $\Delta(\tau_i'-\tau_j')$ with
  \begin{equation} \label{eq:rule}
    \begin{multlined}
    \Delta(\tau_i'-\tau_j') = \sum_{\omega_l \leq 0}
    \frac{\wh{\Delta}_l}{K_l^-(0)} K_l^+(\tau_i')
    K_l^-(\tau_j')  \\ + \sum_{\omega_l > 0} \frac{\wh{\Delta}_l}{(K_l^+(0))^{i-j-1}}
    K_l^+(\tau_i'-\tau_{i-1}') K_l^+(\tau_{i-1}'-\tau_{i-2}') \\ \cdots
    K_l^+(\tau_{j+1}'-\tau_j').
    \end{multlined}
  \end{equation}
  If this procedure is followed for all such hybridization lines, the resulting expression can be rearranged into sums over nested sequences of
  products and convolutions.
The hybridization line connecting to time zero (e.g., $\Delta_1$ in the example above) is excluded because the
corresponding hybridization function only depends on a single time variable, and therefore does not break the convolutional structure of the backbone.

Let us analyze the cost of this procedure. We ignore the $\OO{r}$ cost of
products, since the $\OO{r^2}$ cost of convolutions dominates. Each
hybridization line which is decomposed yields a sum over $r$ frequencies
$\omega_l$, so we obtain a sum over $r^{M-1}$ terms. Each such term contains one
convolution for each of the $2M-2$ internal time variables, yielding an
$\OO{(2M-2) \, r^2}$ complexity per term, or an $\OO{(2M-2) \, r^{M+1}} = \OO{(2M-2) \, (\log (\Lambda) \log (1/\varepsilon))^{M+1}}$ complexity in total.

We note that a similar procedure can be applied to the weak coupling expansion,
with minor modifications. This is described in detail in Appendix
\ref{AppSec:Weak}.

\begin{remark}
  Although we use the DLR expansion to decompose the hybridization functions,
  this is not strictly necessary. Rather, one could expand each hybridization
  function as an arbitrary sum of exponentials, $\Delta(\tau) \approx \sum_{l=1}^p
  K(\tau,\omega_l^\Delta) \wh{\Delta}_l$, tailored to $\Delta$ so that $p < r$,
  and apply the same scheme. This would yield the improved complexity
  $\OO{(2M-2) r^2 p^{M-1}}$. Formulated in the Matsubara frequency domain, this
  gives a rational approximation problem which has been studied for a variety
  of applications in many-body physics, and several approaches have
  been proposed \cite{georges1996,mejutazaera20,shinaoka21,huang23}. We
  use the DLR expansion in the present work for simplicity, and will revisit the problem of a more optimal sum-of-exponentials expansion in future work.
\end{remark}

\subsection{Example: OCA diagram for single-particle Green's function} \label{sec:examples}

To further illustrate the general procedure, we consider the OCA diagram for the single-particle Green's function, which takes the form
\begin{equation} \label{eq:ocacirc}
  \begin{multlined}
  G(\tau) = \int_\tau^\beta d\tau_2 \, \Delta(\tau_2-\tau_1) \mcG_4(\beta-\tau_2)
\mcG_3(\tau_2-\tau) \\ \times \int_0^\tau \, d\tau_1 \, \mcG_2(\tau-\tau_1) \mcG_1(\tau_1).
  \end{multlined}
\end{equation}
For simplicity, we suppress the trace appearing in the single-particle Green's function diagrams, e.g., in \eqref{eq:spgf2}.
In the notation of \eqref{eq:rule}, we have $\tau'_1 = \tau_1$, $\tau'_2 =
\tau$, and $\tau'_3 = \tau_2$. Separating variables in $\Delta(\tau_2-\tau_1)$
and using the identity $K(\tau,\omega) = K(\beta-\tau,-\omega)$, we obtain
\begin{equation} \label{eq:ocacircsep}
\begin{multlined}
  G(\tau) =
    \sum_{\omega_l \leq 0} \frac{\wh{\Delta}_l}{K_l^-(0)} \int_\tau^\beta
    d\tau_2 \, (\mcG_4 K_l^-)(\beta-\tau_2) \mcG_3(\tau_2-\tau) \\ \times \int_0^\tau
    d\tau_1 \, \mcG_2(\tau-\tau_1) (\mcG_1 K_l^-)(\tau_1) \\ + \sum_{\omega_l >
    0} \frac{\wh{\Delta}_l}{K_l^+(0)} \int_\tau^\beta d\tau_2 \,
    \mcG_4(\beta-\tau_2) (\mcG_3 K_l^+)(\tau_2-\tau) \\ \times \int_0^\tau d\tau_1 \,
    (\mcG_2 K_l^+)(\tau-\tau_1) \mcG_1(\tau_1).
  \end{multlined}
\end{equation}
Time-ordered convolutions of the form $\int_\tau^\beta d\tau' \, f(\beta-\tau') g(\tau'-\tau)$ can
be reduced to the standard form introduced above by a change of variables and a reflection
operation, as described in Appendix \ref{app:fastconv}.

A final example for a third-order pseudo-particle self-energy diagram is given in Appendix \ref{app:3rdord}.

\section{Diagrammatic formulation of the algorithm} \label{sec:diagrammatic}

Our procedure can be expressed diagrammatically, which significantly simplifies its implementation. From \eqref{eq:rule}, we see that the terms $\omega_l
\leq 0$ can be expressed by replacing each hybridization line by
a line connecting $\tau = \tau_i$ and $\tau = 0$, labeled by
$K_l^+$, and a line connecting $\tau = 0$ and $\tau = \tau_j$, labeled by $K_l^-$. The terms $\omega_l > 0$ can be expressed by replacing each hybridization line by a chain of lines; one connecting $\tau =
\tau_i$ to $\tau = \tau_{i-1}$, one connecting $\tau = \tau_{i-1}$ to
$\tau = \tau_{i-2}$, and so on, all labeled by $K_l^+$.
For the OCA diagram \eqref{eq:ocasimple}, for example, we obtain
\begin{equation}
  \begin{multlined}
  \Sigma(\tau) = \includegraphics[valign=c,width=4.8cm]{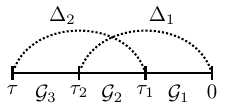} \\
  = \sum_{\omega_l \leq 0} \frac{\wh{\Delta}_{2l}}{K_l^-(0)} \, \,
\includegraphics[valign=c,width=4.8cm]{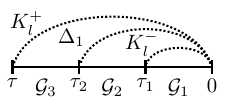} \\ + \sum_{\omega_l > 0}
\frac{\wh{\Delta}_{2l}}{K_l^+(0)} \, \,
\includegraphics[valign=c,width=4.8cm]{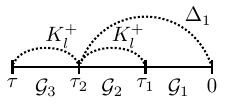}
  \end{multlined}
\end{equation}
which reproduces \eqref{eq:ocasep2}.

This diagrammatic notation can be simplified by observing that lines connecting to
$\tau = 0$ represent a multiplication rather than a convolution, and that all
lines connecting adjacent time variables can be absorbed into the
backbone line connecting those time variables. The above can therefore be replaced by the shorthand
\begin{equation}
  \begin{multlined}
    \Sigma(\tau) = \sum_{\omega_l \leq 0} \frac{\wh{\Delta}_{2l}}{K_l^-(0)} \,\,\,
\includegraphics[valign=c,width=4.8cm]{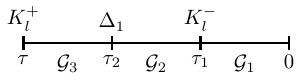}
 \\ + \sum_{\omega_l > 0}
\frac{\wh{\Delta}_{2l}}{K_l^+(0)} \,\,\,
\includegraphics[valign=c,width=4.8cm]{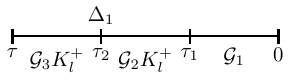},
  \end{multlined}
\end{equation}
where vertical lines centered at a given time variable represent multiplication by the
indicated function, and the functions attached to backbone lines have been
suitably modified. This shorthand notation emphasizes the central idea of our
algorithm, that diagrams can be reduced to sums over backbone
diagrams with a simple convolutional structure.

Using this shorthand, the single-particle Green's function OCA diagram \eqref{eq:ocacirc} is decomposed as
\begin{equation}
  \begin{multlined}
    G(\tau) = \includegraphics[valign=c,width=5.5cm]{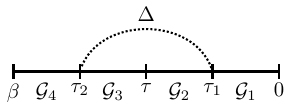} \\
    = \sum_{\omega_l \leq 0} \frac{\wh{\Delta}_l}{K_l^-(0)} \,\,\,
  \includegraphics[valign=c,width=5.5cm]{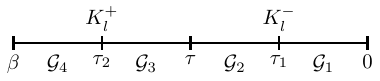} 
  \\ + \sum_{\omega_l > 0} \frac{\wh{\Delta}_l}{K_l^+(0)} \,\,\,
  \includegraphics[valign=c,width=5.5cm]{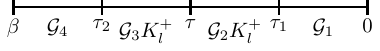},
  \end{multlined}
\end{equation}
which reproduces \eqref{eq:ocacircsep} upon use of the identity $K(\tau,\omega) =
K(\beta-\tau,-\omega)$.

The diagrammatic procedure is illustrated for a third-order self-energy diagram in Appendix \ref{app:3rdord}.

\section{Numerical examples}

We demonstrate an implementation of our algorithm in a strong coupling expansion solver including all self-energy and single-particle Green's function diagrams up to third-order. While our procedure can be applied to diagrams of arbitrary order, in our calculations we have constructed all decompositions by hand, limiting the order in practice. This is, however, a technical limitation, which can be overcome by a code implementing the procedure described in Sections \ref{sec:algorithm} and \ref{sec:diagrammatic} in an automated manner.
All calculations used \texttt{libdlr} for the implementation of the DLR \cite{kaye22_libdlr,libdlr}.

We apply our solver to benchmark systems for which the continuous time hybridization expansion quantum Monte Carlo method (CT-HYB) \cite{werner2006,werner2006a,Haule:2007ys,gull2011} exhibits a severe sign problem \cite{eidelstein2020,zhang2017,zhang2016} due to non-zero off-diagonal hybridization or off-diagonal local hopping in the impurity model.

\subsection{Fermion dimer}

To establish the correctness and order-by-order convergence of our strong coupling solver, we begin by solving impurity models with discrete, finite baths. These systems can be diagonalized exactly, yielding a numerically exact reference for the single-particle Green's function.
We first consider a two-orbital spinless model with inter-orbital hopping, coupled to a discrete bath with off-diagonal hybridization. This minimal model was also used as a benchmark in Ref.~\onlinecite{eidelstein2020}. Its Hamiltonian has the form
\begin{equation} \label{eq:fermion_dimer}
  \begin{multlined}
  H =
  U c_0^\dagger c_0 c^\dagger_1 c_1
  - v \Big( c_0^\dagger c_1 + c_1^\dagger c_0 \Big)
  \\- t \sum_{k=0}^1 \sum_{i=0}^1 \Big( c^\dagger_i b_{ik} + \text{h.c.} \Big)
  - t' \sum_{k=0}^1 \Big( b_{0k}^\dagger b_{1k} + \text{h.c.} \Big),
  \end{multlined}
\end{equation}
where $c_{i}$ is the annihilation operator for the impurity states ($i \in \{0,1\}$) and $b_{ik}$ is the annihilation operator for the bath states $k\in\{0, 1\}$ coupled to the $i$th impurity state. $U$ is the impurity interaction parameter, $v$ is the inter-orbital hopping parameter, $t$ is the parameter for the direct hopping between the impurity and bath orbitals, and $t'$ is the parameter for the inter-bath hopping, which generates an off-diagonal hybridization. 

\begin{figure}
\includegraphics{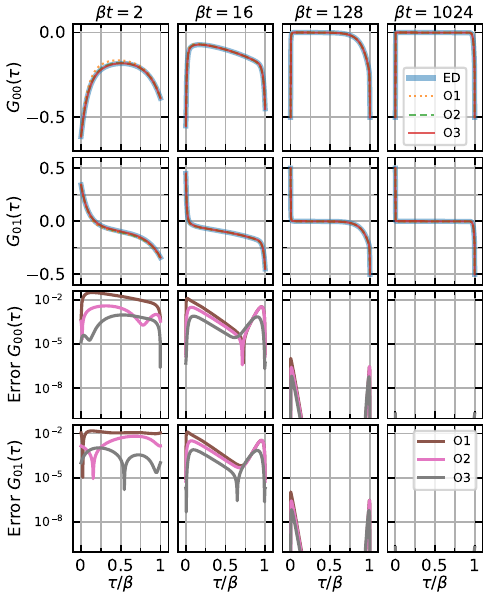}
\caption{Single-particle Green's function for the spinless fermion dimer model \eqref{eq:fermion_dimer}, at inverse temperatures $\beta t = 2$, $16$, $128$, $1024$~(columns), and increasing expansion orders.
The exact diagonal (first row) and off-diagonal (second row) Green's functions obtained from exact diagonalization (ED) are quantitatively described by the first-order approximation (O1).
The diagonal (third row) and off-diagonal (last row) Green's function error decreases when increasing to second (O2) and third (O3) order.
}\label{Fig:numerics_spinless}
\end{figure}

Following Ref.~\onlinecite{eidelstein2020}, we use the parameters $t=1$, $U = 4t$, $v = 3t/2$, $t' = 3t/2$.
In Fig.~\ref{Fig:numerics_spinless}, we compare our strong coupling expansion results for the diagonal ($G_{00}$) and off-diagonal ($G_{01}$) single-particle Green's function at first-, second-, and third-order to the exact solution, for
$\beta = 2$, $16$, $128$, and $1024$. We use the DLR
parameters $\Lambda = 20\beta$ and $\varepsilon=10^{-12}$, yielding $26$, $47$, $71$, and $93$ basis functions,
respectively.
The pseudo-particle self-consistency is iterated until the maximum absolute change in $\mathcal{G}(\tau)$ is less than $10^{-9}$, requiring fewer than 12 iterations (with higher temperatures exhibiting slower convergence). At second-order there are 64 pseudo-particle self-energy and 32 single-particle Green's function diagrams. At third-order there are 2048 pseudo-particle self-energy diagrams (of which 896 are non-zero), and 1024 single-particle Green's function diagrams (of which 448 are non-zero). These numbers account for diagram topologies, hybridization insertions, and forward/backward propagation. The diagram evaluations are independent, enabling perfect parallel scaling.

The error shows an order-by-order convergence, and a rapid
decrease as the temperature is lowered.
The decrease in the error with temperature is a consequence of the ``freezing-out'' of the discrete bath degrees of freedom.
These results demonstrate that a direct diagram evaluation approach is useful for systems in the strong coupling limit even when limited to third-order.
At $\beta=2$ with $r=26$ our third-order calculations required fewer than 0.2 core-hours (for 9 self-consistent iterations), and at $\beta = 1024$ with $r=93$ the same calculation took 4.4 core-hours (for 2 iterations).
These timings can be compared with the 500 core-hours reported for the inchworm Monte Carlo method in Ref.~\onlinecite{eidelstein2020} for the same system (for $\beta = 4 - 64$), though the errors in our third-order calculations are smaller than the stochastic noise shown there in Fig.~1.

\subsection{Two-band Anderson impurity model}

The two-band Anderson impurity model is relevant for the description of correlated $e_g$ bands in transition metal systems. The local Coulomb interaction has the Kanamori form \cite{Kanamori:1963aa}
\begin{equation}
  \begin{multlined}
    H_\text{loc}=U\sum_{\kappa\in\{0,1\}} n_{\kappa\downarrow} n_{\kappa\uparrow} +\sum_{\kappa<\lambda,\sigma} (U'-J_H \delta_{\sigma\sigma'}) n_{\kappa\sigma} n_{\lambda\sigma'} \\ + J_H \sum_{\kappa<\lambda} (c^{\dagger}_{\kappa\uparrow} c^{\dagger}_{\kappa\downarrow} c_{\lambda\downarrow} c_{\lambda\uparrow} + c^{\dagger}_{\kappa\uparrow} c^{\dagger}_{\lambda\downarrow} c_{\kappa\downarrow} c_{\lambda\uparrow}),
  \end{multlined}
\end{equation}
with Hubbard interaction $U$, $U'=U-2J_H$, and Hund's coupling $J_H$.
The Hund's coupling favors high-spin states, and has an important effect on the ordered phases~\cite{hoshino2016} and the dynamics of orbital and spin moments~\cite{werner2008}.
To enable comparison with exact results, we follow Ref.~\onlinecite{eidelstein2020} and consider the two-band impurity model coupled to a bath with an off-diagonal hybridization given by $\Delta_{\kappa\lambda}(\omega) = [\delta_{\kappa\lambda} - s (1 - \delta_{\kappa\lambda})] t^2 \wt{\mathcal{G}}(\omega)$, where $s$ controls the off-diagonal coupling.
We consider two cases: a discrete bath, and a bath with a continuous semicircular spectral function.

\begin{figure}
\includegraphics{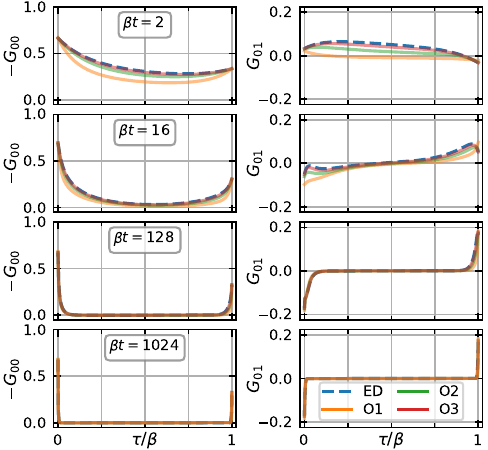}
\caption{Single-particle Green's function for the two-band $e_g$ model with a discrete bath. The left column shows the diagonal Green's function $G_{00}(\tau)$, and the right column shows the off-diagonal component $G_{01}(\tau)$, with decreasing temperatures $\beta t = 2$, $16$, $128$, and $1024$ along the rows. The strong coupling results converge to the exact diagonalization (ED) solution as the expansion order increases.}\label{Fig:numerics_eg}
\end{figure}

In the first case we use a single bath site per orbital. The hybridization function is given by  $\wt{\mathcal{G}}(\omega) = \sum_k \delta(\omega - \epsilon_k)$ with $\epsilon_k \in \{ \pm 2.3t \}$, and we use a strong off-diagonal hybridization $s=1/2$.
This case can be solved using exact diagonalization, and provides a non-trivial test case.
The rapid order-by-order convergence of the orbitally-resolved Green's function $G_{\lambda\kappa}(\tau-\tau')=-\langle c_{\lambda}(\tau) c_{\kappa}^{\dagger}(\tau') \rangle$ computed by our strong coupling solver is shown in Fig.~\ref{Fig:numerics_eg}
at $\beta = 2$, $16$, $128$, and $1024$.
We use the DLR parameters $\Lambda = 12.5 \beta$ and $\varepsilon = 10^{-8}$, yielding $17$, $32$, $46$, and $61$ basis functions, respectively. The pseudo-particle self-consistency is iterated until the maximum absolute change in $\mathcal{G}(\tau)$ is less than $10^{-6}$, requiring fewer than 12 iterations for the temperatures and expansion orders considered. At second-order there are 256 pseudo-particle self-energy and 64 single-particle Green's function diagrams. At third-order there are 16384 pseudo-particle self-energy diagrams (of which 14080 are non-zero), and 4096 single-particle Green's function diagrams (of which 3520 are non-zero).

As in the fermion dimer example, the error decreases with the temperature.
Interestingly, the results show that the off-diagonal component $G_{01}(\tau)$ becomes substantially enhanced at lower temperatures, with sharp features emerging around $\tau=0$ and $\tau=\beta$. Physically, these features correspond to short-time quantum fluctuations between orbitals, and should be important for the stabilization of orbital orders.
The DLR discretization of the Green's function is able to capture such features significantly more efficiently than a standard equispaced grid discretization.

\begin{figure}
\includegraphics{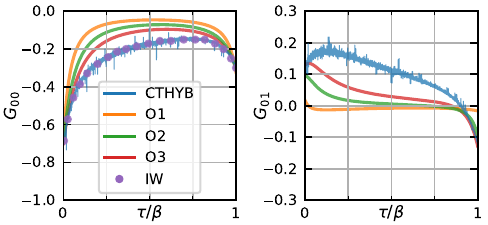}
\caption{Single-particle Green's function for the two-band $e_g$ model with a metallic bath at inverse temperature $\beta t = 8$. The diagonal part G$_{00}$ is shown in the left panel (a), and the off-diagonal part G$_{01}$ in the right panel (b). The strong coupling expansion converges towards the result produced by the continuous time hybridization expansion method (CT-HYB) and the inchworm quantum Monte Carlo (IW) results from Ref.~\onlinecite{eidelstein2020}, but in this case diagrams beyond third-order contribute significantly.}\label{Fig:numerics_eg_bethe}
\end{figure}

We next present results for the metallic semicircular hybridization function given by $\wt{\mathcal{G}}(\omega) = \frac{2}{\pi D^2}\sqrt{D^2 - \omega^2}$, with $s = 1$ and $D = 2t$, at inverse temperature $\beta = 8/t$. We use the DLR parameters $\lambda = 10\beta$ and $\varepsilon = 10^{-10}$, yielding $25$ basis functions. The pseudo-particle self consistency converges below $10^{-6}$ in fewer than 30 iterations.
In Fig.~\ref{Fig:numerics_eg_bethe}, we compare our approach with
inchworm Monte Carlo \cite{cohen2015taming, antipov2017, eidelstein2020}
and continuous time hybridization expansion Monte Carlo (CT-HYB) \cite{gull2011,werner2006,werner2006a,Haule:2007ys},
using the TRIQS \cite{Parcollet2015398} implementation \cite{TRIQSCTHYB} for the latter.
For CT-HYB the average sign is approximately $0.25$, and reducing the variance is costly, though techniques such as improved estimators \cite{PhysRevB.85.205106, PhysRevB.94.125153, PhysRevB.100.075119} and worm sampling \cite{PhysRevB.92.155102, WALLERBERGER2019388} could in principle mitigate this difficulty. However, the exponential decay of the sign with temperature severely restricts this method.
The inchworm algorithm does not suffer the same kind of sign problem \cite{cohen2015taming, eidelstein2020}, and inchworm Monte Carlo results down to $t\beta = 64$ have been reported \cite{eidelstein2020}.

While our strong coupling solver still converges with the order, the third-order solution differs significantly from the exact solution.
This system lies outside the reach of a third-order strong coupling expansion, as is expected since the model is far from the strong coupling limit, and would require the inclusion of higher-order diagrams.
We note, however, that our result was obtained at a significantly lower cost than the corresponding inchworm calculation (154 core-hours vs.\ 1,500 core-hours at $t\beta = 8$), so including higher-order diagrams within our framework should be tested for comparison.

\subsection{Minimal model for \texorpdfstring{Ca$_2$RuO$_4$}{Ca2RuO4}}

Our previous results suggest that the strong coupling expansion converges rapidly in the insulating regime. For many insulating systems, like Mott insulators with tetragonal symmetry and strong spin-orbit coupling, off-diagonal hybridization plays an important role. This represents a substantial challenge for Monte-Carlo based solvers, due to the sign problem. 
The most promising workaround was introduced in Ref.~\onlinecite{zhang2016}, which used the interaction expansion in combination with a basis rotation to reduce the sign problem and enable an analysis of a spin-orbit coupled system at elevated temperatures. We have seen that the strong coupling expansion proposed in this work is easily extendable to spin-orbit coupled systems as long as the system is deep within the Mott insulating phase. Here, we provide a proof-of-principle calculation for a minimal model of Ca$_2$RuO$_4$~\cite{zhang2017,gorelov2010,sutter2017}.

The electronic configuration of Ca$_2$RuO$_4$ includes four electrons in the three t$_{2g}$ orbitals which at low temperatures undergo an isosymmetric structural transition 
accompanied by a Mott metal-insulator transition. This structural distortion reduces the energy of the d$_{xy}$ orbital, making it doubly occupied. The remaining orbitals with two electrons undergo a Mott metal-insulator transition, leading to the $S=1$ state~\cite{liebsch2007,sutter2017,gorelov2010}. In the t$_{2g}$ space, the matrix representation of the orbital moment operators is (up to a sign) equal to the $L=1$ operator in the cubic basis, an observation which is often called TP correspondence~\cite{sugano2012}. An open question is the nature of the magnetic moments due to strong spin-orbit coupling. Two scenarios were proposed in the literature: (i) the spin-orbit coupling leads to a correction of the $S=1$ picture and induces a single-ion anisotropy~\cite{zhang2017}, and (ii) the spin-orbit coupling changes the moment of the ground state to j$_{\text{eff}}=0$~\cite{khaliullin2013,akbari2014}.
 It is difficult to distinguish these two scenarios a priori from the value of the spin-orbit coupling, as its effect can be substantially enhanced due to a dynamical increase of the spin-orbit effect. Answering these questions therefore requires unbiased simulations. Our goal is not to solve the question of Ca$_2$RuO$_4$, but rather to show that on the level of the minimal model, we can capture the competition between all relevant interactions. We leave the extension of our approach to full ab-initio models, and the resolution of the question of magnetism in Ca$_2$RuO$_4$, as an important future problem.

We consider a three-orbital Hubbard model spanned by d$_{xy}$, d$_{xz}$ and d$_{yz}$ orbitals within the DMFT approximation, which maps the lattice problem to an impurity problem. The local part of the impurity problem is given by
\begin{equation}
  \begin{multlined}
    H_{\text{loc}}=
    H_{\text{LS}} +
    \sum_{\substack{\kappa\sigma}}  \left[\epsilon_{\kappa}-\mu \right] c_{\kappa\sigma}^{\dagger} c_{\kappa\sigma} + U \sum_{\kappa} n_{\kappa \uparrow} n_{\kappa \downarrow} 
 \\ + \sum_{\kappa < \lambda} \sum_{\sigma, \sigma'}
   (U' - J_H\delta_{\sigma \sigma'}) n_{\kappa \sigma} n_{\lambda \sigma'} \\
   + J_H \sum_{\kappa < \lambda} 
 \left(
 c^{\dagger}_{\kappa \uparrow} c^{\dagger}_{\kappa \downarrow} c_{\lambda \downarrow} c_{\lambda \uparrow}
  +c^{\dagger}_{\kappa \uparrow} c^{\dagger}_{\lambda \downarrow} c_{\kappa \downarrow} c_{\lambda \uparrow} 
 \right),
  \end{multlined}
\end{equation}
where $\kappa, \lambda \in \{d_{xy},d_{xz},d_{yz}\}$,
and the on-site energies $\epsilon_{xz}=\epsilon_{yz}=0$ are split with respect to the doubly occupied d$_{xy}$ orbital by the crystal field $\epsilon_{xy}=\Delta_{\text{cf}} = -0.5\,$eV.
We choose the chemical potential $\mu$ such that the system is occupied by four electrons on average. The interacting part of the Hamiltonian is given by the Slater-Kanamori interaction parametrized by the Hubbard interaction $U$ and the Hund's coupling $J_H$. We use the established values $U=2.3$ eV and $J_H=0.4$ eV obtained from the constrained random phase approximation~\cite{sutter2017,zhang2017,mravlje2011}.
The spin-orbit coupling introduces a complex coupling between the t$_{2g}$ orbitals. By employing the TP correspondence we obtain
\begin{equation}
  H_{\text{LS}}=\lambda_{\text{SOC}} \vec{L} \cdot \vec{S} =\frac{\I \lambda_{\text{SOC}}}{2}\sum_{\kappa\lambda\mu,\sigma\sigma'}\epsilon_{\kappa\lambda\mu}\tau^{\mu}_{\sigma\sigma'} c_{\kappa\sigma}^{\dagger} c_{\lambda\sigma'},
\end{equation}
where $\lambda_{\text{SOC}} = 0.1\,$eV is the size of the spin-orbit coupling,  $\epsilon$ is the Levi-Civita matrix element, and $\tau^{\nu}$ is the $\nu$th Pauli matrix. 

\begin{figure}
\includegraphics{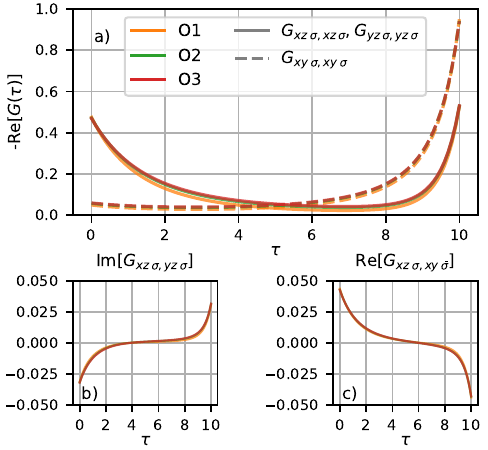}
\caption{
  (a) Diagonal component of the single-particle Green's function, with increasing expansion orders, for the three-band effective model of Ca$_2$RuO$_4$. G$_{xz \, \sigma, xz \, \sigma }(\tau)$ and $G_{yz \, \sigma, yz \, \sigma}(\tau)$ are degenerate and close to half-filling (solid lines), while $G_{xy \, \sigma, xy \, \sigma}(\tau)$ is close to unity-filling (dashed lines). (b) Imaginary part of the off-diagonal component $G_{xz \, \sigma, yz \, \sigma}(\tau)$. (c) Real part of the off-diagonal component $G_{xz \, \sigma, xy \bar{\sigma}}(\tau)$ generated by the local spin-orbit coupling. Calculations were performed at inverse temperature $\beta=10\,$eV$^{-1}$.}
\label{Fig:cro_soc}
\end{figure}

We solve the problem on the Bethe lattice, for which the DMFT self-consistency condition is particularly simple. The hybridization function $\Delta_{\lambda\kappa}$ is obtained from the local Green's function $G_{\lambda\kappa}(\tau-\tau')=-\langle \mathcal{T} c_{\lambda}(\tau) c_{\kappa}^{\dagger}(\tau') \rangle$  as $\Delta_{\lambda\kappa}=t_{\lambda} G_{\lambda\kappa}t_{\kappa}.$ We restrict to intra-orbital transitions given by $t_{\lambda}=\{t_{d_{xy}},t_{d_{xz}},t_{d_{yz}}\}$, and the hopping integrals are estimated as $t_{d_{yz}}=t_{d_{xz}}=0.25\,$eV and $t_{d_{xy}}=0.5\,$eV to match the bandwidth measured by ARPES and previous theoretical studies~\cite{sutter2017}. All calculations are performed at inverse temperature $\beta=10$ eV$^{-1}.$

To the authors' knowledge, ours is the first self-consistent third-order strong coupling DMFT calculation for a three-band model.
We use the DLR parameters $\Lambda = 100$ and $\varepsilon = 10^{-8}$, yielding $26$ basis functions. We perform the pseudo-particle self-energy and DMFT lattice self-consistencies in tandem, with a tolerance threshold of $10^{-6}$ for changes in the respective propagators, requiring 11 iterations for the first-order calculation, 9 iterations for the second-order calculation, and 7 iterations for the third-order calculation. At each order we use the solution from the previous order for the initial iterate.
The calculation took approximately 90,000 core-hours (11 hours on 8,192 cores) using our preliminary code, with the independent evaluation of 186,624 third-order diagrams performed in parallel.
Although the use of the DLR enables efficient diagram evaluation at very low temperatures, we find that the convergence of the DMFT and pseudo-particle self-consistency exhibits a critical slow-down as the temperature is lowered. We attribute this to the presence of the antiferromagnetic instability of the two half-filled orbitals, as observed in Ca$_2$RuO$_4$, which becomes antiferromagnetic at the N\'eel temperature $T_N \approx 110\,$K \cite{PhysRevB.58.847}.

We plot the diagonal part of the single-particle propagator $G_{\alpha\alpha}$ for $\alpha\in\{d_{xy},d_{xz},d_{yz}\}$ in Fig.~\ref{Fig:cro_soc}(a). We observe that the d$_{xy}$ orbital is almost fully occupied, and the d$_{xz}$ and d$_{yz}$ orbitals are half-filled due to the strong Coulomb interaction. In agreement with the previous examples, we observe rapid convergence with increasing diagram order. We observe a maximum absolute difference between the second- and third-order calculations of less than $6 \times 10^{-3}$, which gives a reasonable estimate of the error in the second-order calculation. Thus, the third-order calculation not only gives us a more accurate result than this, but allows us to estimate the error of the second-order calculation. The main effect of the higher-order diagrams is to enhance the value of the propagators away from the edges of the interval $[0,\beta]$, which we interpret as an increase in charge fluctuations with the increasing order of the expansion.
These results allow us to estimate the orbital polarization $p=n_{xy}-(n_{xz}+n_{yz})/2$, which in the system with finite spin-orbit coupling is given by $p=0.84$, and is similar to the result without the spin-orbit coupling, $p_{\lambda=0}=0.87$. This is consistent with the conclusion of Ref.~\onlinecite{zhang2017}, in which the persistence of strong orbital polarization upon inclusion of the spin-orbit coupling was used as a signature of the $S=1$ picture. While the spin-flip and pair-hopping terms were neglected in Ref.~\onlinecite{zhang2017}, our results confirm that at the level of the minimal model, their effect would not produce a significant qualitative change.

Due to the spin-orbit coupling, the single-particle propagators $G_{\lambda\kappa}$ obtain complex off-diagonal components, as shown in Fig.~\ref{Fig:cro_soc}(b) for $\Im G_{xz \, \sigma, yz \, \sigma}(\tau)$ and in Fig.~\ref{Fig:cro_soc}(c) for $\Re G_{xz \, \sigma, xy \, \bar{\sigma}}(\tau)$.
Hence, the spin-orbit coupling also generates a dynamic mixing of spin and orbitals, coupling the out-of-plane orbitals $xz$ and $yz$ for equal spin, and the in-plane orbital $xy$ with the spin-flipped component of the out-of-plane orbitals.
This can be understood from the structure of the spin-orbit coupling in the t$_{2g}$ subspace of the d-orbital cubic harmonics~\cite{stamokostas2018}.
It is the sign change in these off-diagonal components as a function of $\tau$ which generates a dynamical sign problem in the hybridization determinant of the CT-HYB algorithm \cite{gull2011,werner2006,werner2006a,Haule:2007ys}.

Application of our method to realistic materials beyond our simplified proof-of-principle model would require extension to a given lattice structure, and combination with first-principles calculations, both well-established techniques. This will be explored in our future work.

\section{Conclusion} \label{sec:conclusion}

We have proposed and implemented a new algorithm
for the fast evaluation of imaginary time Feynman diagrams. By taking
advantage of a separability property of imaginary time objects, the algorithm obtains
a decomposition which can be evaluated efficiently within the DLR basis. We have
developed the method in detail for the bold strong coupling expansion of the
Anderson impurity model, and showcase an implementation up to third-order. Its
application to the weak coupling expansion is described in Appendix
\ref{AppSec:Weak}. 
The extension to higher-order diagrams is
straightforward and will be pursued in future work.
A parallel implementation of our algorithm provides a path towards robust and
high-order accurate evaluation of diagrams at least up to intermediate orders at very low temperature,
with no Monte Carlo integration and no sign problem.
The combination of our approach with methods for diagrammatic expansions of very high-order, including
Monte Carlo~\cite{gull2010,gull2011} and TCI~\cite{fernandez22,erpenbeck2023}, is a topic of our current research. In particular, TCI might be used to exploit compressibility across imaginary time and orbital degrees of freedom, while maintaining the robustness, high-order accuracy, and favorable scaling at low temperatures of our scheme. In general, we expect the ideas presented here to serve as useful tools for future algorithmic development.

We envision that the ideal short term applications of the proposed method are
multi-orbital systems within the Mott insulating regime involving strong off-diagonal
hybridization terms induced by either spin-orbit coupling or symmetry-broken
phases. Examples include Ca$_2$RuO$_4$ \cite{zhang2017,gorelov2010,sutter2017}, Sr$_2$IrO$_4$~\cite{jackeli2009,kim2009,lenz2019,kim2012}, and Nb$_3$Cl$_8$ \cite{grytsiuk2023nb3cl8}. In this limit a reliable description is obtained by a relatively low-order expansion, but these systems are still challenging for current state-of-the-art Monte Carlo techniques due to the sign problem. A particularly appealing direction is to enter the symmetry-broken phase and study dynamical properties of exotic magnetic phases such as canted (anti)ferromagnetism~\cite{kim2012}. The robustness, speed, and low memory footprint of our algorithm should allow it to couple well with ab-initio descriptions based on DFT+DMFT in Mott insulators, as implemented in existing numerical libraries, e.g. TRIQS/DFTTools \cite{aichorn16}. 

An efficient automated implementation of our algorithm, allowing for the evaluation of diagrams of arbitrary order and topology, is under development. Improvements to the algorithm, involving further decomposition of diagrams as well as more efficient representations of the hybridization function, are also being explored, and are expected to yield significant further reductions in computational cost. Beyond imaginary time diagrams, we believe the idea of fast diagram evaluation through compression of the integrand, either through separation of variables or otherwise, represents a promising research frontier with the potential to circumvent many of the limitations of traditional schemes. 

\acknowledgments

We thank O. Parcollet, A. Georges, R. Rossi, and G. Cohen for helpful discussions.
Z.H. is supported by the Simons Investigator Award, which is a grant from the Simons Foundation (825053, Lin Lin).
H.U.R.S. acknowledges funding from the European Research Council (ERC) under the European Union’s Horizon 2020 research and innovation programme (Grant Agreement No.\ 854843-FASTCORR).
Computations were enabled by resources provided by the National Academic Infrastructure for Supercomputing in Sweden (NAISS) and the Swedish National Infrastructure for Computing (SNIC)
through projects
SNIC 2022/21-15, 
SNIC 2022/13-9, 
SNIC 2022/6-113, 
and 
SNIC 2022/1-18, 
at PDC, NSC, and CSC, partially funded by the Swedish Research Council through Grant Agreements No.\ 2022-06725 and No.\ 2018-05973.
D.G. is supported by the Slovenian Research Agency (ARRS) under Programs J1-2455, MN-0016-106 and P1-0044.
The Flatiron Institute is a division of the Simons Foundation.

\appendix

\section{Fast convolution of DLR expansions}
\label{app:fastconv}

In Ref.~\onlinecite{kaye22_dlr}, the imaginary time convolution
\begin{equation} 
  H(\tau) = \int_0^\beta d \tau' \, F(\tau - \tau') G(\tau')
  \label{eq:tau_convolution}
\end{equation}
is expressed in terms of the contraction of the vectors of values of $F$
and $G$ at the DLR nodes $\tau_k$ with a rank three tensor:
$H(\tau_i) = \sum_{j,k=1}^r \mathcal{C}_{ijk} F(\tau_j)
  G(\tau_k)$.
One can similarly write the time-ordered convolution
\begin{equation} \label{eq:ppsc_convolution} 
  H(\tau) =
  \int_0^\tau d \tau' \,
  F(\tau - \tau') G(\tau')
\end{equation}
as a tensor contraction.
The cost of this approach scales as $O(r^3)$.
We demonstrate that these convolutions can be computed in only $\OO{r^2}$
operations by writing the action of the convolution tensor on the DLR
coefficients directly. We focus on \eqref{eq:ppsc_convolution}, but the method
for \eqref{eq:tau_convolution} is analogous. 

Using the explicit formula \eqref{eq:kernel}, we first compute the time-ordered convolution \eqref{eq:ppsc_convolution} of
$K(\tau,\omega)$ and $K(\tau,\omega')$:
\begin{equation}
  \begin{multlined}
  \int_0^\tau d\tau' \, K(\tau - \tau', \omega) K(\tau', \omega')
  \\=
  \begin{cases}
  \frac{ K(0, \omega) K(\tau, \omega') - K(\tau, \omega) K(0, \omega') }{ \omega
    - \omega' } \, , & \omega \ne \omega' \\
  \tau K(0, \omega) K(\tau, \omega) \, , & \omega = \omega'.
  \end{cases}
\end{multlined}
\end{equation}
Given the DLR expansions $F(\tau) = \sum_{j=1}^r K(\tau,\omega_j)
\wh{f_j}$ and
$G(\tau) = \sum_{k=1}^r K(\tau,\omega_k) \wh{g_k}$, with possibly matrix-valued $\wh{f_j}$ and $\wh{g_k}$, this yields 
\begin{widetext}
\begin{equation} \label{eq:convolution_F1} 
\begin{aligned}
  H(\tau) &= \sum_{j,k=1}^r \wh{f_j} \wh{g_k} \int_0^\tau d\tau' \, K(\tau - \tau', \omega_j) K(\tau', \omega_k)
\\ &=
  \tau \sum_{j=1}^r K(\tau, \omega_j) K(0, \omega_j) \wh{f_j} \wh{g_j}
  +
  \sum_{j=1}^r K(\tau, \omega_j)
  \left(
    \paren{\sum_{\substack{k=1 \\ k\ne j}}^r    
    \frac{K(0, \omega_k)}{\omega_k - \omega_j} \wh{f_k}} \wh{g_j} 
    +
    \wh{f_j}
    \sum_{\substack{k=1 \\ k\ne j}}^r    
    \frac{K(0, \omega_k)}{\omega_k - \omega_j} \wh{g_k}
    \right)
\equiv
  \tau H_1(\tau) + H_2(\tau).
\end{aligned}
\end{equation}
\end{widetext}
Here, we recognize $H_1$ and $H_2$ as DLR expansions themselves.
Thus, $H$ can be obtained at the DLR nodes $\tau_k$ by computing the DLR
coefficients of $H_1$ and $H_2$ directly from those of $F$ and $G$, at an
$\OO{r^2}$ cost, and
then evaluating their DLR expansions to obtain the values $H(\tau_k) = \tau_k
H_1(\tau_k) + H_2(\tau_k)$. We note that if one wishes to
obtain the DLR coefficients of $H$ directly, then it is slightly more efficient
to precompute the matrix of multiplication by $\tau$ in its DLR coefficient
representation, and apply this directly to the computed vector of DLR
coefficients of $H_1$. Adding the result to the vector of DLR
coefficients of $H_2$ yields that of $H$.

Since
\begin{equation}
  \begin{multlined}
  \int_\tau^\beta d\tau' F(\beta-\tau') G(\tau'-\tau) \\ = \int_0^{\beta-\tau} d\tau' F(\beta-\tau-\tau') G(\tau') = H(\beta-\tau)
  \end{multlined}
\end{equation}
for $H(\tau) \equiv \int_0^\tau d\tau' F(\tau-\tau') G(\tau')$,
convolutions of this form appearing in the single-particle Green's function diagrams can be reduced to the form \eqref{eq:ppsc_convolution}. One can therefore compute $H(\tau)$ by the method described above and a reflection $H(\tau) \mapsto H(\beta-\tau)$, a linear map which can be represented in the DLR basis.

We lastly mention that \eqref{eq:tau_convolution} is given by
\begin{equation}\label{Eq:dlr_convolution}
  \begin{multlined}
  H(\tau) = 
  \sum_{j=1}^r K(\tau, \omega_j) (\tau - K(1,\omega_j)) \wh{f_j} \wh{g_j}
  +
  \sum_{j=1}^r K(\tau, \omega_j) \\ \times
  \left(
    \paren{\sum_{\substack{k=1 \\ k\ne j}}^r    
    \frac{\wh{f_k}}{\omega_k - \omega_j}} \wh{g_j} 
    +
    \wh{f_j}
    \sum_{\substack{k=1 \\ k\ne j}}^r    
    \frac{\wh{g_k}}{\omega_k - \omega_j} \right).
  \end{multlined}
\end{equation}

\section{Decomposition of third-order pseudo-particle self-energy diagram}
\label{app:3rdord}

We illustrate the decomposition procedure described in Section \ref{sec:procedure} for a third-order self-energy diagram:
\begin{widetext}
\begin{equation} \label{eq:3rdordsep}
  \begin{aligned}
    \Sigma(\tau) &= 
    \begin{multlined}[t]
      \int_0^\tau d\tau_4 \, \int_0^{\tau_4} d\tau_3 \, \int_0^{\tau_3}
    d\tau_2 \, \int_0^{\tau_2} d\tau_1 \Delta_3(\tau-\tau_1)
    \Delta_2(\tau_4-\tau_2) \Delta_1(\tau_3) \\ \times \mcG_5(\tau-\tau_4)
    \mcG_4(\tau_4-\tau_3) \mcG_3(\tau_3-\tau_2) \mcG_2(\tau_2-\tau_1)
    \mcG_1(\tau_1)
    \end{multlined} \\
    &= 
    \begin{multlined}[t]
    \sum_{\omega_k,\omega_l \leq 0} \frac{\wh{\Delta}_{3k}
    \wh{\Delta}_{2l}}{K_k^-(0) K_l^-(0)} K_k^+(\tau) \int_0^\tau d\tau_4 \,
    \mcG_5(\tau-\tau_4) K_l^+(\tau_4) \int_0^{\tau_4} d\tau_3 \, \mcG_4(\tau_4-\tau_3)
    \Delta_1(\tau_3) \\
    \times \int_0^{\tau_3} d\tau_2 \, \mcG_3(\tau_3-\tau_2)
    K_l^-(\tau_2)
    \int_0^{\tau_2} d\tau_1 \,  \mcG_2(\tau_2-\tau_1) (\mcG_1 K_k^-)(\tau_1) \\
    + \sum_{\omega_k \leq 0, \omega_l > 0} \frac{\wh{\Delta}_{3k}
    \wh{\Delta}_{2l}}{K_k^-(0) K_l^+(0)}
     K_k^+(\tau) \int_0^\tau d\tau_4 \,
    \mcG_5(\tau-\tau_4) \int_0^{\tau_4} d\tau_3 \, (\mcG_4 K_l^+)(\tau_4-\tau_3)
    \Delta_1(\tau_3) \\
    \times \int_0^{\tau_3} d\tau_2 \, (\mcG_3 K_l^+)(\tau_3-\tau_2)
    \int_0^{\tau_2} d\tau_1 \,  \mcG_2(\tau_2-\tau_1) (\mcG_1 K_k^-)(\tau_1) \\
    + \sum_{\omega_k > 0, \omega_l \leq 0} \frac{\wh{\Delta}_{3k}
    \wh{\Delta}_{2l}}{(K_k^+(0))^3 K_l^-(0)} \int_0^\tau d\tau_4
    \, (\mcG_5 K_k^+)(\tau-\tau_4) K_l^+(\tau_4) \int_0^{\tau_4} d\tau_3 \,
    (\mcG_4 K_k^+)(\tau_4-\tau_3) \Delta_1(\tau_3) \\ \times \int_0^{\tau_3}
    d\tau_2 \, (\mcG_3 K_k^+)(\tau_3-\tau_2) K_l^-(\tau_2) \int_0^{\tau_2}
    d\tau_1 \,  (\mcG_2 K_k^+)(\tau_2-\tau_1) \mcG_1(\tau_1) \\
    + \sum_{\omega_k, \omega_l  > 0} \frac{\wh{\Delta}_{3k}
    \wh{\Delta}_{2l}}{(K_k^+(0))^3 K_l^+(0)} 
    \int_0^\tau d\tau_4
    \, (\mcG_5 K_k^+)(\tau-\tau_4) \int_0^{\tau_4} d\tau_3 \,
    (\mcG_4 K_k^+ K_l^+)(\tau_4-\tau_3) \Delta_1(\tau_3) \\ \times \int_0^{\tau_3}
    d\tau_2 \, (\mcG_3 K_k^+ K_l^+)(\tau_3-\tau_2) \int_0^{\tau_2}
    d\tau_1 \,  (\mcG_2 K_k^+)(\tau_2-\tau_1) \mcG_1(\tau_1).
  \end{multlined}
\end{aligned}
\end{equation}
\end{widetext}
Here we have introduced the notation
$(\mcG_i K_k^\pm K_l^\pm)(\tau) \equiv \mcG_i(\tau) K(\tau,\pm\omega_k)
K(\tau,\pm \omega_l)$.
To obtain this result diagrammatically using the procedure described in Section \ref{sec:diagrammatic}, we proceed step by step. We first decompose
$\Delta_3(\tau-\tau_1)$:
\begin{widetext}
\begin{equation}
  \begin{multlined}
  \Sigma(\tau) = \includegraphics[valign=c,width=6.2cm]{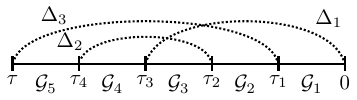} \\
  = \sum_{\omega_k \leq 0} \frac{\wh{\Delta}_{3k}}{K_k^-(0)} \, \,
\includegraphics[valign=c,width=6.0cm]{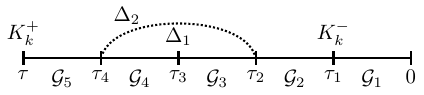} + \sum_{\omega_k > 0}
\frac{\wh{\Delta}_{3k}}{(K_k^+(0))^3} \,\,
  \includegraphics[valign=c,width=6.0cm]{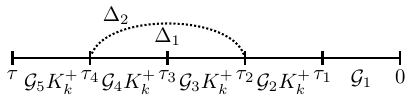}.
  \end{multlined}
\end{equation}
\end{widetext}
Then, decomposing $\Delta_2(\tau_4-\tau_2)$ in both of the resulting diagrams reproduces \eqref{eq:3rdordsep}:
\begin{widetext}
\begin{equation}
  \begin{multlined}
    \Sigma(\tau) = \sum_{\omega_k,\omega_l \leq 0} \frac{\wh{\Delta}_{3k}
    \wh{\Delta}_{2l}}{K_k^-(0) K_l^-(0)} \,\,\,
    \includegraphics[valign=c,width=8.0cm]{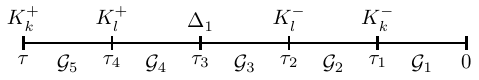} \\
    + \sum_{\omega_k \leq 0, \omega_l > 0} \frac{\wh{\Delta}_{3k}
    \wh{\Delta}_{2l}}{K_k^-(0) K_l^+(0)} \,\,\,
     \includegraphics[valign=c,width=8.0cm]{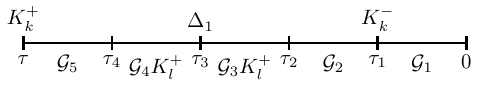} \\
    + \sum_{\omega_k > 0, \omega_l \leq 0} \frac{\wh{\Delta}_{3k}
    \wh{\Delta}_{2l}}{(K_k^+(0))^3 K_l^-(0)} \,\,\,
    \includegraphics[valign=c,width=8.0cm]{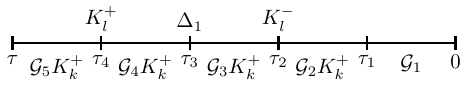} \\
    + \sum_{\omega_k, \omega_l  > 0} \frac{\wh{\Delta}_{3k}
    \wh{\Delta}_{2l}}{(K_k^+(0))^3 K_l^+(0)} \,\,\,
    \includegraphics[valign=c,width=8.0cm]{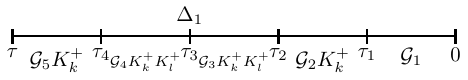}.
  \end{multlined}
\end{equation}
\end{widetext}

\section{Weak coupling interaction expansion}\label{AppSec:Weak}
In this Appendix, we describe the application of our algorithm to the interaction
expansion. We write the non-interacting part of the action using the quadratic term in \eqref{Eq:Ham} and the hybridization function in \eqref{Eq:Act_hyb}: 
\begin{equation}
  \begin{multlined}
  \mathcal{S}_0 = \sum_{\kappa,\lambda=1}^n
  \int_0^\beta d\tau \, c^\dagger_\lambda(\tau) \left[ \partial_{\tau} -\epsilon_{\kappa\lambda} -\Delta_{\lambda\kappa}(\tau-\tau')\right]c_\kappa(\tau')
  \end{multlined}.
\end{equation}
We then carry out an expansion in the interacting part of the action, given by
\begin{equation}
\begin{aligned}
\mathcal{S}_{\text{int}}&=\int_0^\beta d\tau \, H_{\text{int}} \\
&= \int_0^\beta d\tau \, \sum_{\kappa,\lambda,\mu,\nu=1}^n U_{\kappa\lambda\mu\nu} c^{\dagger}_{\kappa}(\tau) c^{\dagger}_{\lambda}(\tau) c_{\nu}(\tau)  c_{\mu}(\tau),
\end{aligned}
\end{equation}
which yields an expression for the interacting propagator, 
\begin{equation}
\label{Eq:Expansion}
\begin{multlined}
G_{\kappa\lambda}(\tau)=\frac{1}{Z}\sum_{m=0}^{\infty} \int_0^{\beta} d\tau_1\cdots d\tau_m \\ \times \langle \mathcal{T} H_{\text{int}}(\tau_1) \ldots H_{\text{int}}(\tau_m) c_{\kappa}^{\dagger}(\tau) c_{\lambda}(0) \rangle_{\mathcal{S}_0},
\end{multlined}
\end{equation}
for $\lambda, \kappa = 1,\ldots,n$.
Here $\mathcal{T}$ 
is the time-ordering operator and the expectation value is evaluated with respect to the non-interacting part of the action: $\langle \cdot \rangle_{\mathcal{S}_0}=\Tr[e^{-\mathcal{S}_0} \, \cdot]$. By the linked cluster theorem, all disconnected diagrams in the expansion cancel with the partition function $Z$, and we need only consider connected diagrams.
The expectation value in \eqref{Eq:Expansion} can be evaluated using Wick's theorem, and the non-interacting propagator is given by the Weiss Green's function $\mathcal{G}_0$, which satisfies the Dyson equation $\mathcal{G}_{0}=g_0+g_0\ostar \Delta \ostar \mathcal{G}_{0}$.
Here, the convolution is defined as
\[(a\ostar b)_{jk}(\tau)=\sum_{l=1}^n \int_0^{\beta} d\tau' \, a_{jl}(\tau - \tau') b_{lk}(\tau'),\]
and the non-interacting propagator 
is given by $g_0(\tau) = K(\tau,\epsilon-\mu)$, with $\epsilon$ the single-particle Hamiltonian (quadratic part of $H_{\text{loc}}$), $K$ defined by \eqref{eq:kernel}, and $\mu$ the chemical potential.

The propagator $G(\tau)$ is obtained by first evaluating the self-energy $\Sigma(\tau)$ and then solving the Dyson equation $G=\mathcal{G}_0+\mathcal{G}_0\ostar\Sigma\ostar G.$ One can represent the self-energy either using a bare expansion, in which diagrams depend on the bare propagator $\mathcal{G}_0$, or a bold expansion, in which diagrams depend on the full propagator $G$. For concreteness, we focus on the case of the bare expansion, but the procedure described below is consistent with both schemes. We note, however, that the bold scheme requires solving the Dyson equation self-consistently.

The first-order terms of the bare expansion are given by the Hartree and Fock diagrams, which depend only on the single-particle density matrix. The first retarded diagrams appear at second-order, with a representative example given by  
\begin{equation}
  \Sigma_{\alpha\beta}(\tau)=-U_{\beta'\gamma\beta\gamma'} U^{\alpha\delta\alpha'\delta'}  \mathcal{G}_{0,\alpha'\beta'}(\tau) \mathcal{G}_{0,\gamma'\delta}(-\tau) \mathcal{G}_{0,\delta'\gamma}(\tau).
\end{equation}
Here, we use the Einstein notation that all repeated indices are summed over. In general, the second-order diagrams for the weak coupling expansion are similar to the NCA diagrams for the strong coupling expansion, in that they involve only multiplications in imaginary time.

The first diagrams involving imaginary time integration appear at third-order. A representative example is 
\begin{equation}
\begin{multlined}
  \Sigma_{\alpha\beta}(\tau)=-U_{\beta'\delta\beta\delta'} U_{\gamma\epsilon\gamma'\epsilon'} U_{\alpha\omega\alpha'\omega'}  \mathcal{G}_{0,\delta'\omega}(-\tau)\\ \times \int_0^{\beta} d\tau_1
  \mathcal{G}_{0,\alpha'\gamma}(\tau-\tau_1) \mathcal{G}_{0,\omega'\epsilon}(\tau-\tau_1) \\ \times \mathcal{G}_{0,\gamma'\beta'}(\tau_1) \mathcal{G}_{0,\epsilon'\delta}(\tau_1).
\end{multlined}
\end{equation}
We see that the third-order weak coupling diagrams have a simple convolutional structure,
precisely of the form \eqref{eq:tau_convolution}, and the DLR-based method described in Appendix \ref{app:fastconv} can be directly used to evaluate them. We refer the reader to Ref.~\onlinecite{tsuji2013nonequilibrium} for the expressions for the other third-order diagrams, which have a similar structure.

\begin{remark}
We pause to consider the summation over orbital indices, which we have thus far assumed is carried out explicitly. For multi-orbital systems, this leads to a large number of terms, growing exponentially with the diagram order. Methods based on sparsity or decomposability of tensors can in some cases be used to handle this issue, and one must verify their compatibility with our approach to imaginary time integration. If the interaction tensor is sparse, the orbital index sums can be taken over a subset of terms, which is evidently compatible with our approach. This is the case in many physically interesting settings: for example, in cubic crystals in which $d$ orbitals are split into $e_g$ and $t_{2g}$ manifolds, the Coulomb integral is given by the Slater-Kanamori interaction, and for the $t_{2g}$ subspace the interaction tensor has only $21$ out of $81$ non-zero entries~\cite{Kanamori:1963aa,nilsson2017}. More sophisticated schemes, such as Cholesky decomposition \cite{beebe77}, density fitting \cite{weigend09,ren12}, tensor hypercontraction \cite{lu15,yeh23}, and the canonical polyadic decomposition \cite{udo11,pierce21,pierce22} aim to decompose the interaction tensor using a low-rank structure, but these are typically applied in quantum chemistry or real materials calculations in which the orbital index dimension is significantly larger than that considered here. Although schemes of this type would likely also be compatible with our approach, further research is needed to determine which, if any, would be appropriate, and this question is outside the scope of the present work. Thus, in the remainder of this Appendix, we consider explicit summation over orbital indices (or, in the sparse case, a subset of them), and for simplicity focus on each term separately. Reverting to the notation used in the description of our scheme for strong coupling diagrams, we refer to different components of the Weiss field $\mathcal{G}_0$ as $\mathcal{G}_1$, $\mathcal{G}_2$, etc, in lieu of orbital indices, notating that unlike in the strong coupling case, each $\mathcal{G}_k$ is a scalar-valued function.
\end{remark}

\begin{figure}
\includegraphics[width=1.0\linewidth]{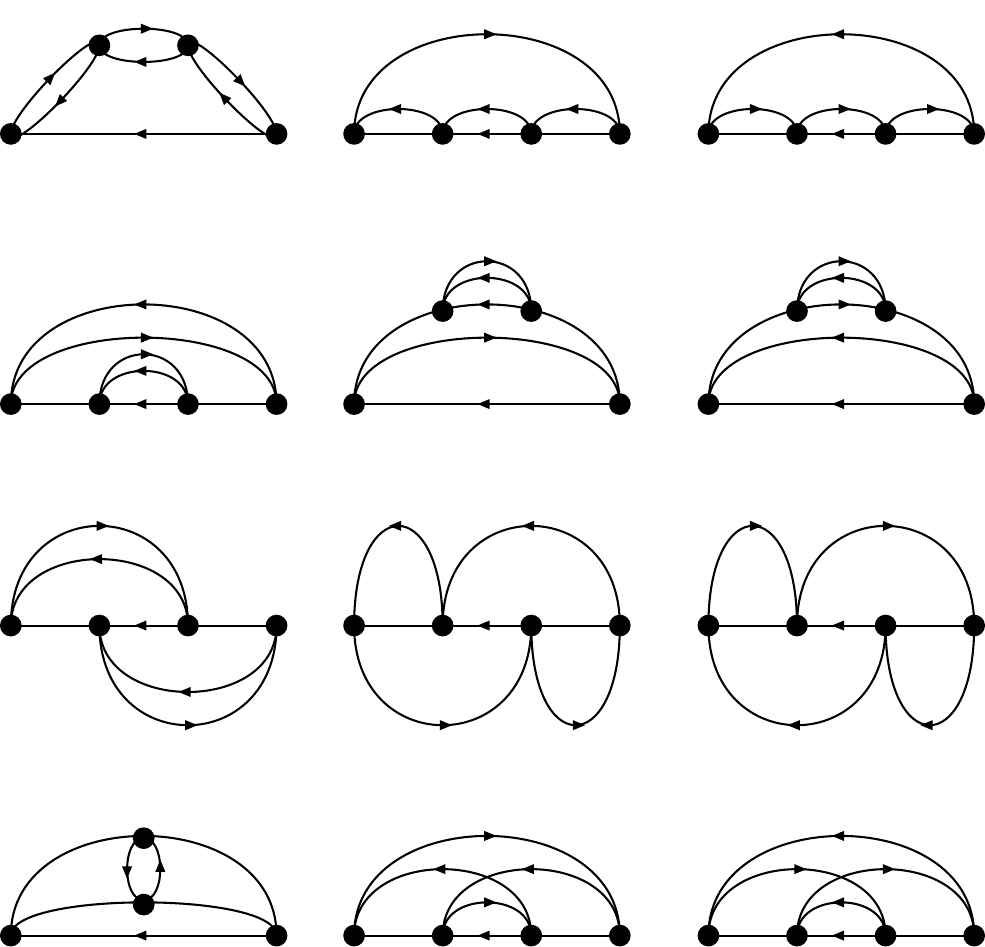}
\caption{Fourth-order self-energy diagrams for the weak coupling expansion of the Anderson impurity problem assuming a paramagnetic phase at half filling. Lines represent the electronic propagator $\mathcal{G}_0~(G)$ for the bare~(bold) expansion, and dots correspond to the interaction vertices $U$.}
\label{Fig:sigma_weak}
\end{figure}
 
The first non-convolutional diagrams appear at fourth-order. In total, there are 12 topologically distinct fourth-order diagrams, shown in Fig.~\ref{Fig:sigma_weak}. As for the third-order diagrams, the diagrams in the first two rows can be written as nested sequences of products and convolutions~\cite{tsuji2013nonequilibrium}, and evaluated using our efficient convolution scheme. For example, each orbital index combination of the first diagram in the first row of Fig.~\ref{Fig:sigma_weak} takes the form
\begin{equation}
\begin{multlined}
  \Sigma(\tau) = \mathcal{G}_7(\tau)\int_0^{\beta} d\tau_2 \int_0^{\beta} d\tau_1 \mathcal{G}_6(\tau-\tau_1) \\ \times
   \mathcal{G}_5(\tau-\tau_1) \mathcal{G}_4(\tau_1-\tau_2) \mathcal{G}_3(\tau_1-\tau_2)\mathcal{G}_2(\tau_2)\mathcal{G}_1(\tau_2).
\end{multlined}
\end{equation}
We note that the diagrams in the second row are not irreducible, and in the case of the bold expansion would be omitted.

On the other hand, the diagrams in the third and fourth rows of of Fig.~\ref{Fig:sigma_weak} cannot be expressed in terms of nested convolutions, and we must apply our decomposition scheme. For example, for the first diagram in the third row, we obtain expressions of the form
\begin{equation}\label{Eq:fourth_example}
  \begin{multlined}
  \Sigma(\tau) = \int_0^{\beta} d\tau_2 \, \int_0^{\beta} d\tau_1 \,
  \mathcal{G}_7(\tau-\tau_2) \\ \times \mathcal{G}_6(\tau_2) \mathcal{G}_5(-\tau_2) \mathcal{G}_4(\tau_2-\tau_1)  \mathcal{G}_3(\tau_1) \\ \times \mathcal{G}_2(\tau-\tau_1) \mathcal{G}_1(\tau_1-\tau).
  \end{multlined}
\end{equation}
Structurally, this diagram somewhat resembles the strong coupling OCA diagrams, but here the integral is taken over the full square $[0,\beta]^2$ rather than a time-ordered subset. To avoid discontinuities, we divide the integral into six parts, each corresponding to an ordering of the variables $\tau$, $\tau_1$, and $\tau_2$:
\begin{equation*}
  \begin{aligned}
\int_0^\beta d\tau_2 \int_0^\beta d\tau_1 &= 
\int_0^\tau d\tau_2 \int_0^{\tau_2} d\tau_1 + \int_0^\tau d\tau_1 \int_0^{\tau_1} d\tau_2 \\
&+ \int_\tau^\beta d\tau_2 \int_0^\tau d\tau_1 + \int_\tau^\beta d\tau_1 \int_0^\tau d\tau_2 \\ 
&+ \int_\tau^\beta d\tau_2 \int_{\tau_2}^\beta d\tau_1 + \int_\tau^\beta d\tau_1 \int_{\tau_1}^{\beta} d\tau_2.
  \end{aligned}
\end{equation*}
The first two terms are structurally equivalent to the strong coupling OCA self-energy diagrams, the third and fourth to the OCA single-particle Green's function diagrams, and, up to a simple change of variables, the fifth and sixth also to the OCA self-energy diagrams. We can therefore apply the same methodology with only minor modifications, in each case expanding a single function in the DLR basis in order to decompose the diagram. The computational complexity of each such diagram evaluation is therefore the same as for the strong coupling OCA diagrams, with two important differences: (1) six terms must be computed instead of one, and (2) the functions $\mathcal{G}_k$ are scalar-valued, rather than $2^n \times 2^n$ matrix-valued.

It is straightforward to verify that the other fourth-order diagrams in the third and fourth rows of Fig.~\ref{Fig:sigma_weak} have a similar structure: each has a backbone of propagators whose convolutional structure is broken by propagators coupling non-adjacent time variables. At fourth-order, the convolutional structure can be reinstated by separating variables in at most two functions, as in the third-order strong coupling diagrams. All higher-order self-energy diagrams share an analogous pattern, as in the strong coupling case, though as the number of integration variables grows, the integrals must be split into a factorially growing number of time-ordered terms. 

\bibliography{dlrdiag}

\end{document}